\documentclass[prx,twocolumn,english,superscriptaddress,floatfix,longbibliography]{revtex4-2}

\usepackage{graphicx}
\usepackage{dcolumn}
\usepackage{bm}
\usepackage{amsmath}
\usepackage{amssymb}
\usepackage{braket}

\begin{document}

\title[VAE for Learning VQE Circuit Parameters and Quantum Phase Transitions]{Learning Variational Quantum Eigensolver Circuit Parameters with Classical Artificial Intelligence for Quantum Phase Transition Detection}

\author{Xin Li}
    \email{lihsin922@outlook.com}
    \affiliation{Center for Quantum Technology Research and Key Laboratory of Advanced Optoelectronic Quantum Architecture and Measurements (MOE), \\ School of Physics, Beijing Institute of Technology, Beijing 100081, China}
\author{Zhang-Qi Yin}
    \email{zqyin@bit.edu.cn}
    \affiliation{Center for Quantum Technology Research and Key Laboratory of Advanced Optoelectronic Quantum Architecture and Measurements (MOE), \\ School of Physics, Beijing Institute of Technology, Beijing 100081, China}

\date{\today}

\begin{abstract}
Learning many-body quantum states and quantum phase transitions remains a major challenge in quantum many-body physics. Classical machine learning methods offer certain advantages in addressing these difficulties. In this paper, we shift the research perspective that bypasses the need for high-fidelity quantum states by directly learning the parameters of parameterized quantum circuits. By integrating attention mechanisms with a variational autoencoder (VAE), we efficiently capture hidden correlations within the parameter distribution that encode quantum phase boundaries. These correlations enable us to extract phase transition information in an unsupervised manner and identify a data-driven ``generalized order parameter'' from the learned latent space. The framework remains robust even when VQE converges to local minima rather than true ground states. Moreover, our framework acts as a classical representation of VQE optimization landscapes, enabling the reconstruction of complete phase diagrams. Our results demonstrate that quantum phase transition information can indeed be extracted directly from variational circuit parameters.
\end{abstract}


\maketitle

\section{Introduction}

Quantum phase transitions (QPTs) are qualitative changes in the ground state of quantum many-body systems driven by quantum fluctuations at zero temperature~\cite{Sachdev2011, Sachdev_2023}. They underlie fundamental phenomena from symmetry breaking and topological order to high-temperature superconductivity, making their detection and characterization a central challenge in condensed matter physics. The advent of Noisy Intermediate-Scale Quantum (NISQ) devices~\cite{preskill2018quantum} offers unprecedented opportunities for probing ground-state physics and quantum phase transitions through variational quantum algorithms~\cite{Islam_2011, Bloch2012}. In particular, the Variational Quantum Eigensolver (VQE)~\cite{peruzzo2014variational, Kandala2017, Endo2018, McClean_2016, Moll_2018} enables approximate ground state preparation without requiring fault-tolerant error correction, opening new pathways for studying QPTs on near-term quantum hardware~\cite{TILLY20221, Cerezo2021VQA}.

However, detecting and characterizing QPTs remains a formidable challenge, especially for topological quantum phase transitions where conventional local order parameters fail~\cite{PhysRevLett.101.010504, Pollmann_2012, PhysRevLett.109.050402}. These transitions are characterized by nonlocal many-body topological invariants (MBTI) or entanglement properties~\cite{PhysRevB.81.064439, Elben_2020, Dehghani_2021}, which from a quantum computing perspective require either full quantum state tomography or reconstruction of nonlocal reduced density matrices---both prohibitively expensive for NISQ devices.

The rapid development of machine learning and quantum computation has spurred numerous approaches to this problem. Machine learning methods have been introduced to explore quantum phase transitions: manifold learning has been applied to analyze Hamiltonian parameter spaces~\cite{Lidiak_2020}, CNNs have been used to learn from quantum Monte Carlo results~\cite{Dong_2019}, and supervised learning has been employed to extract phase information from experimental data~\cite{Rem2019}, etc. Quantum-native approaches such as Quantum Convolutional Neural Networks (QCNNs)~\cite{Cong_2019, Herrmann_2022, Monaco_2023, chen2025quantumensemblelearningprogrammable, sander2024quantumconvolutionalneuralnetwork} and classical shadow tomography~\cite{Huang_2020, Elben_2020, Huang_2022, Lewis_2024} enable efficient phase detection by leveraging specific quantum measurements or symmetry properties. Unsupervised learning techniques have been successfully applied to quantum many-body snapshots~\cite{Broecker_2017, PhysRevB.100.045129, Rahaman_2023, kasatkin2024classifimunsupervisedmethoddetect, Ziv2025unsupervised, Marashli2025identifying, Chowdhury2024autoencoder}, while VQE-based methods~\cite{Lively2024noise, Cao2025unveiling, Angelides_2025, PhysRevResearch.5.043217, PhysRevE.107.024113} extract phase information from carefully chosen observables.

Despite these advances, a fundamental limitation persists: existing methods typically require access to high-fidelity ground states and prior knowledge of appropriate observables or symmetries. In VQE implementations, achieving such fidelity is often precluded by barren plateaus~\cite{McClean2018, PRXQuantum.3.010313, PRXQuantum.2.040316, Wang_2021} and local minima trapping~\cite{anschuetz2022quantum, chen2023localminimaquantumsystems, apte2022nonconvexoptimizationhamiltonianalternation, li2025quantum}, which are ubiquitous in complex many-body energy landscapes. Nevertheless, intriguing recent work suggests that phase information may remain extractable even from local minima~\cite{cao2024unveilingquantumphasetransitions, crognaletti2024equivariant}, hinting that valuable physics is encoded beyond the quantum state itself.

Motivated by these observations, in this work we shift the analytical focus from quantum state properties to VQE-converged circuit parameters. To validate this approach, we construct a classical deep neural network that learns directly from the distribution of VQE-optimized parameters, thereby circumventing stringent requirements on quantum state fidelity and establishing a purely classical paradigm requiring no additional quantum resources. We apply this machine learning framework to study QPTs in different models (Transverse-Field Ising Model and Cluster-Ising Model) using various VQE ansätze. Our results demonstrate that the framework successfully extracts phase transition information for both Landau-type transitions characterized by local order parameters and topological quantum phase transitions. Moreover, through the learned representations we identify a data-driven ``generalized order parameter.'' Crucially, our numerical results show that this approach remains effective even when VQE converges to local minima rather than true ground states, providing a robust pathway for QPT detection in the NISQ era. Our work demonstrates that quantum phase transition information can indeed be extracted directly from VQE circuit parameters from the VQE convergence process.

The remainder of this paper is organized as follows. Section~\ref{sec:motivation} presents the physical motivation and problem formulation. Section~\ref{sec:framework} details our classical deep neural network framework combining attention mechanisms with variational autoencoders. Section~\ref{sec:results} presents comprehensive numerical validation on the Transverse-Field Ising Model and Cluster-Ising model. Section~\ref{sec:discussion} discusses the physical interpretation, robustness to VQE imperfections, and broader implications. Section~\ref{sec:conclusion} concludes.

\section{Motivation and Problem Formulation}
\label{sec:motivation}

Consider a quantum many-body system governed by a set of continuous physical parameters $\boldsymbol{x}$, with the Hamiltonian denoted as $H(\boldsymbol{x})$. Here, $\boldsymbol{x}$ represents the physical quantities driving the system through phase transitions. For instance, in the Transverse-Field Ising Model (TFIM), $\boldsymbol{x}$ corresponds to the transverse magnetic field strength $h$; whereas in the XXZ spin chain model, it corresponds to the $ZZ$ interaction strength $\Delta$ and the $XX+YY$ exchange coupling strength $J$. As the control parameter $\boldsymbol{x}$ varies continuously, the ground state $\ket{\psi_0(\boldsymbol{x})}$ of the system may undergo abrupt qualitative changes, signifying a quantum phase transition (QPT). When employing the Variational Quantum Eigensolver (VQE) to approximate the ground states for this family of Hamiltonians $H(\boldsymbol{x})$, we fix a parameterized quantum circuit architecture $U(\boldsymbol{\theta})$ as the ansatz, where $\boldsymbol{\theta} \in \mathbb{R}^N$ denotes the $N$-dimensional tunable circuit parameters. For each given physical configuration $\boldsymbol{x}$, the objective of VQE is to determine the optimal set of parameters $\boldsymbol{\theta}^*(\boldsymbol{x})$ that minimizes the variational energy:
\begin{equation}
    \boldsymbol{\theta}^*(\boldsymbol{x}) = \arg\min_{\boldsymbol{\theta}} \bra{0} U^\dagger(\boldsymbol{\theta}) H(\boldsymbol{x}) U(\boldsymbol{\theta}) \ket{0}.
\end{equation}

\begin{figure}[htbp]
    \centering
    \includegraphics[width=0.5\textwidth]{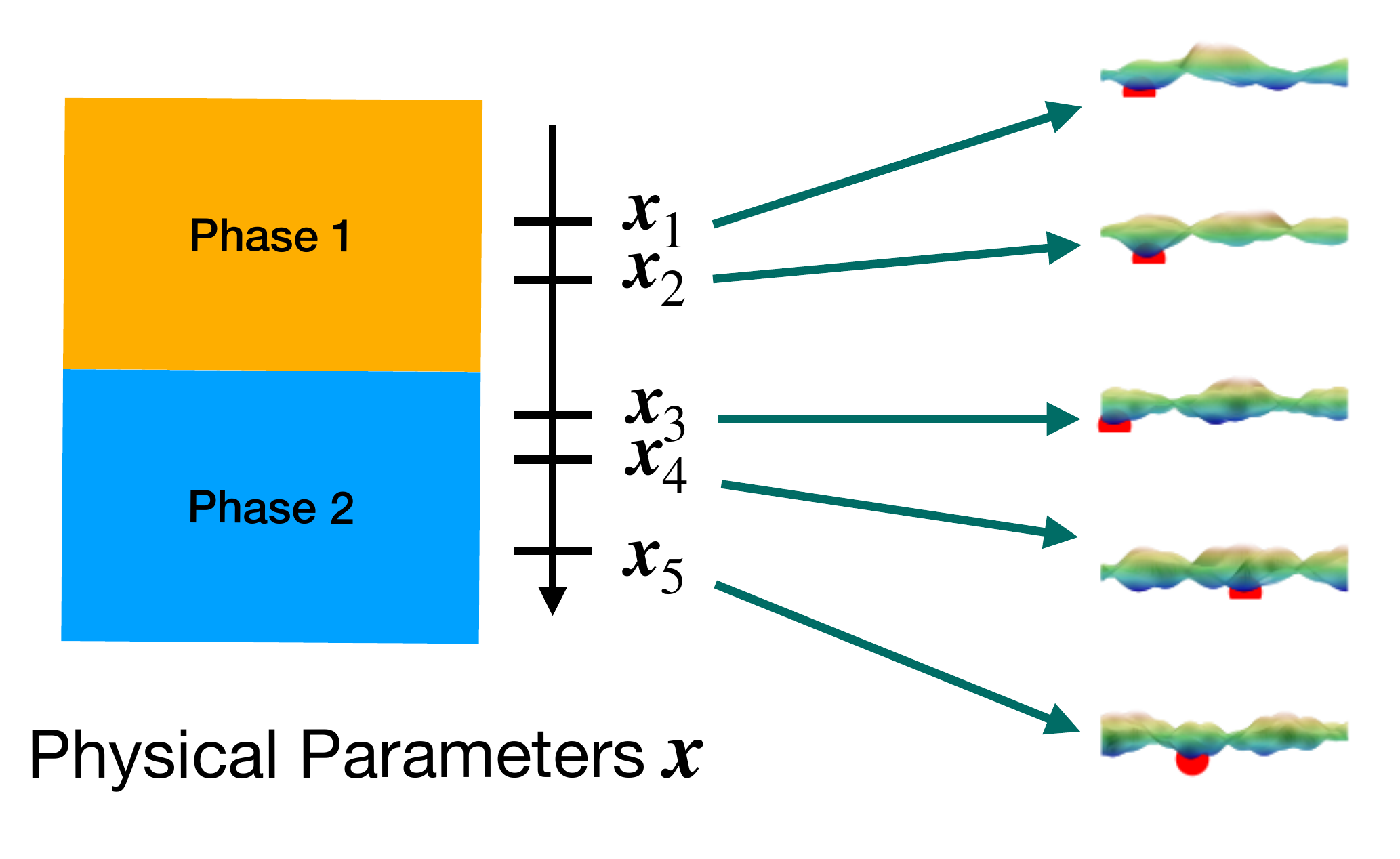}
    \caption{
       Schematic of VQE energy landscapes and endpoints for different parameters $\boldsymbol{x}$. A three-dimensional landscape is used to represent the energy landscape surface in high-dimensional parameter space. Different physical parameters $\boldsymbol{x}$ correspond to local minima under different energy landscapes. The final converged local minima are highlighted with red dots.
    }
    \label{fig:vqe_map}
\end{figure}

To investigate the correlation between the parameter space and quantum phases, we impose a consistent initialization condition across the optimization processes for all $\boldsymbol{x}$. Specifically, all VQE optimizations are initialized from the same starting point $\boldsymbol{\theta}_{\text{initial}}$. Under this condition, the VQE optimizer effectively defines a non-linear mapping from the physical parameter space to the variational circuit parameter space:
 \begin{equation}\label{eq:vqe_map}
    \mathcal{M}: \boldsymbol{x} \mapsto \boldsymbol{\theta}^*(\boldsymbol{x})
\end{equation}
This implies that for every physical configuration $\boldsymbol{x}$, the optimization process drives the parameters from the common starting point $\boldsymbol{\theta}_{\text{initial}}$, evolving along the gradient flow of the energy landscape, and eventually settling at a specific endpoint $\boldsymbol{\theta}^*$ within the high-dimensional parameter space. Figure~\ref{fig:vqe_map} visualizes this mapping. Under the same ansatz, different physical parameters $\boldsymbol{x}$ generate distinct energy landscapes. Consequently, starting from an identical initialization, the optimization trajectories converge to different endpoints in the circuit parameter space.

Our core premise is as follows: since ground states $\ket{\psi_0(\boldsymbol{x})}$ belonging to different quantum phases possess fundamentally distinct entanglement structures and symmetry properties, the ansatz circuit must be adjusted to distinctly different regions of the parameter space to approximate these divergent wavefunctions. It is reasonable to hypothesize that for physical parameters within the same quantum phase, the corresponding optimal circuit parameters $\boldsymbol{\theta}^*$ should be proximate to one another under a certain metric in the high-dimensional space, thereby forming continuous clusters. Conversely, when $\boldsymbol{x}$ crosses a phase transition point, the abrupt change in the global properties of the ground state wavefunction forces the VQE optimizer to seek a set of parameters with a significantly different structure. This results in observable discontinuities or pattern transitions in the parameter distribution. Consequently, we recast the detection of QPTs as a high-dimensional data mining problem. By uncovering the manifold structure underlying the distribution of endpoints $\{ \boldsymbol{\theta}^*(\boldsymbol{x}) \}$, we employ classical deep neural network models to map these parameters into a latent space capable of automatic clustering. This allows for the identification of quantum phase transitions without the need to measure physical observables.

\begin{figure}[htbp]
    \centering
    \includegraphics[width=0.5\textwidth]{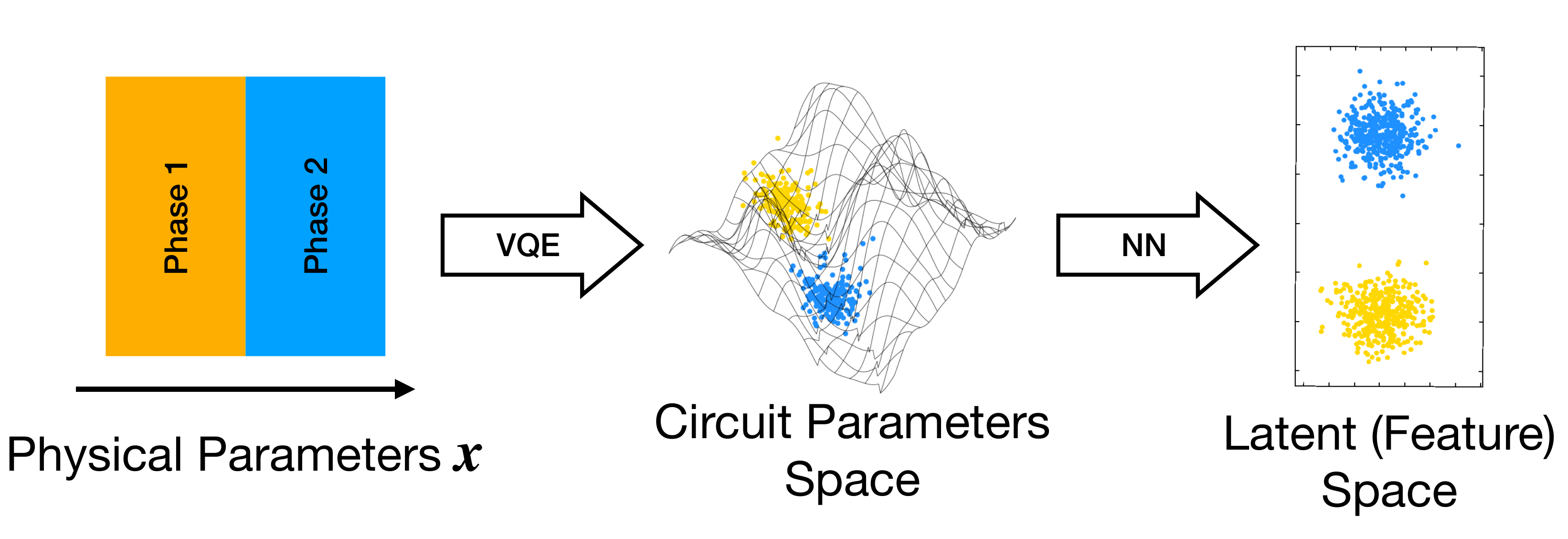}
    \caption{Schematic diagram of the unsupervised quantum phase recognition framework. The arrow VQE indicates VQE optimization under the same initialization for all physical parameters $\boldsymbol{x}$. The arrow NN indicates the training of a classical neural network model on the parameters $\boldsymbol{\theta}^*(\boldsymbol{x})$ obtained from VQE optimization.}
    \label{fig:schematic_uns_learning_mo}
\end{figure}

Figure~\ref{fig:schematic_uns_learning_mo} illustrates the workflow of our unsupervised quantum phase recognition framework. In the physical parameter space formed by all Hamiltonians $\boldsymbol{x}$, distinct regions correspond to different quantum phases. The VQE optimization process can be viewed as a mapping $\mathcal{M}: \boldsymbol{x} \mapsto \boldsymbol{\theta}^*(\boldsymbol{x})$. Starting from a consistent initialization $\boldsymbol{\theta}_{\text{init}}$, physical parameters from different phases drive the circuit parameters to converge into distinct regions within the high-dimensional circuit parameter space. A classical neural network model extracts implicit correlations within the parameter distribution, mapping the high-dimensional parameters to a low-dimensional latent space to achieve automatic clustering and recognition of quantum phases.

\section{Methods}
\label{sec:framework}

The design of our neural architecture is motivated by the specific characteristics of VQE parameter data. Unlike images or time series, VQE circuit parameters $\boldsymbol{\theta}$ inhabit a high-dimensional space where physical correlations are distributed non-locally across parameter indices. Traditional linear dimensionality reduction methods such as PCA~\cite{Jolliffe2016PCAReview, pearson1901liii} or shallow clustering algorithms like Gaussian Mixture Models (GMM)~\cite{1199538, 6790375} are inadequate for capturing these complex dependencies. Deep Embedded Clustering (DEC)~\cite{xie2016unsuperviseddeepembeddingclustering} and its variational variant~\cite{jiang2017variationaldeepembeddingunsupervised} address this by exploiting deep neural networks for feature extraction combined with classical clustering. Recent work has incorporated attention mechanisms into VAEs to enhance representational capabilities~\cite{Correia_2023}. Our architecture builds on these advances while being specifically tailored to the non-local structure of quantum circuit parameters.

\subsection{Attention Mechanism}
\label{sec:attention_mechanism}

The attention mechanism was introduced to address the information bottleneck in neural sequence processing~\cite{bahdanau2016}. The Transformer architecture~\cite{vaswani2023attentionneed} replaced recurrent structures with self-attention, enabling parallel computation and direct modeling of dependencies between arbitrary sequence positions regardless of their separation. This architecture has since revolutionized numerous domains, including computer vision through Vision Transformers~\cite{dosovitskiy2021imageworth16x16words} and diffusion models through transformer-based architectures~\cite{peebles2023scalablediffusionmodelstransformers}.

Formally, given an input sequence $\mathbf{X} = [\mathbf{x}_1, \mathbf{x}_2, \dots, \mathbf{x}_n]$ of length $n$, the self-attention mechanism computes an output sequence $\mathbf{Z} = [\mathbf{z}_1, \mathbf{z}_2, \dots, \mathbf{z}_n]$ where each $\mathbf{z}_i$ aggregates information from all positions through a weighted combination. This is achieved via the Query-Key-Value (QKV) formulation: each input vector $\mathbf{x}_i$ is linearly projected into query $\mathbf{q}_i = \mathbf{x}_i\mathbf{W}^Q$, key $\mathbf{k}_i = \mathbf{x}_i\mathbf{W}^K$, and value $\mathbf{v}_i = \mathbf{x}_i\mathbf{W}^V$ representations.

The attention weights are computed as the normalized similarity between queries and keys:

\begin{equation}
    \text{Attention}(Q, K, V) = \text{softmax}\left(\frac{QK^T}{\sqrt{d_k}}\right)V
    \label{eq:scaled_attention}
\end{equation}

where $d_k$ is the dimensionality of the key vectors and the scaling factor $\sqrt{d_k}$ prevents gradient vanishing~\cite{vaswani2023attentionneed}. The output $\mathbf{z}_i$ is the weighted sum of value vectors $\mathbf{v}_j$ according to these attention weights.

To capture multiple types of dependencies simultaneously, we employ multi-head self-attention (MHSA):

\begin{equation}
    \text{MultiHead}(Q, K, V) = \text{Concat}(\text{head}_1, \dots, \text{head}_h)\mathbf{W}^O
    \label{eq:multihead_attention}
\end{equation}

where $\text{head}_i = \text{Attention}(Q\mathbf{W}_i^Q, K\mathbf{W}_i^K, V\mathbf{W}_i^V)$ and $h$ is the number of attention heads. This allows the model to attend to different representation subspaces in parallel~\cite{vaswani2023attentionneed}.

\subsection{VAE Framework Overview}
\label{subsec:vae_framework}

\begin{figure}[htbp]
    \centering
    \includegraphics[width=0.48\textwidth]{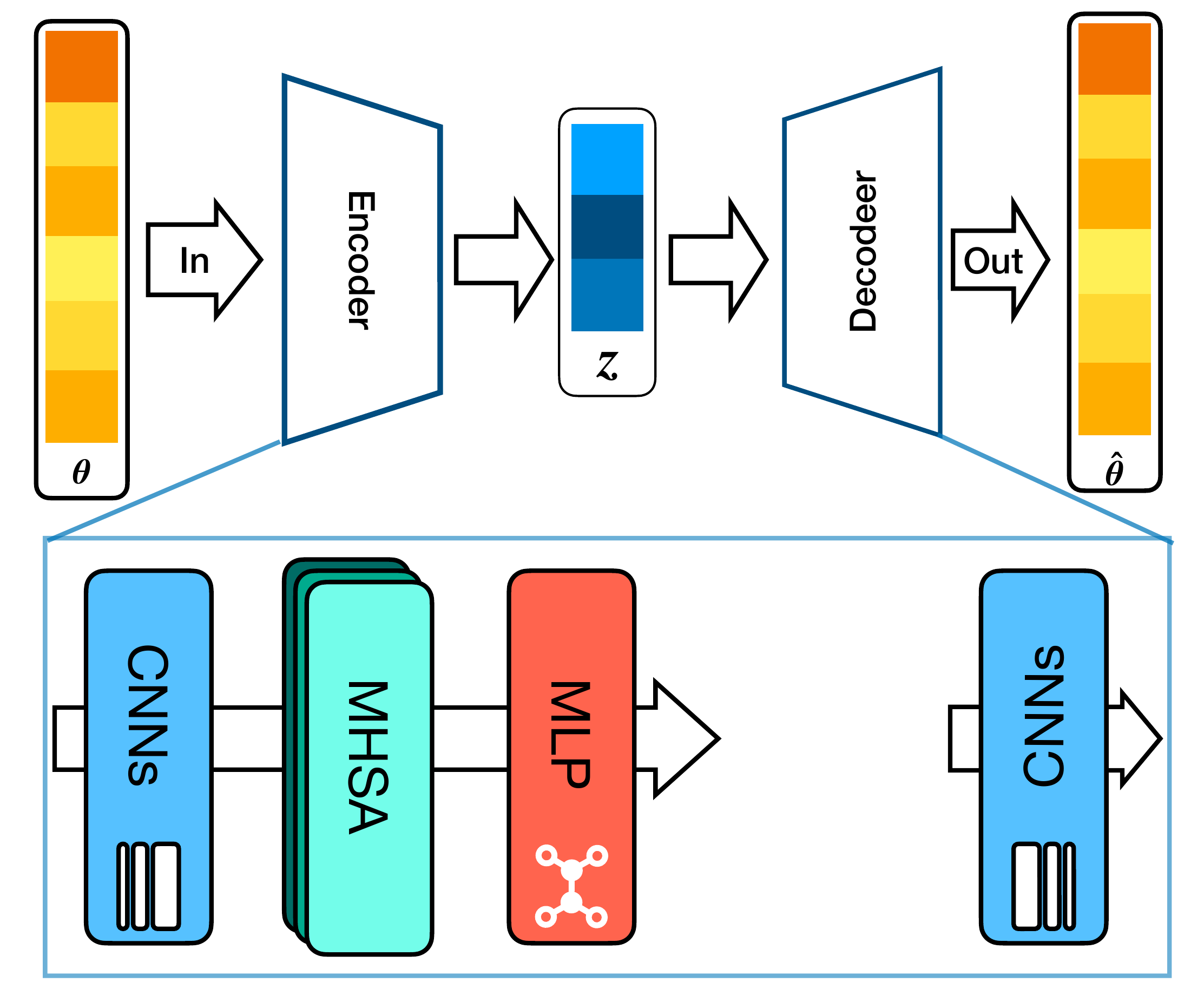}
    \caption{Overview of the VAE framework. The encoder (left) processes input parameters through 1D-CNN, MHSA, and MLP modules to produce the parameters of a latent Gaussian distribution. The decoder (right) reconstructs parameters from latent samples using a mirrored deconvolutional architecture.}
    \label{fig:vae_framework}
\end{figure}

Our VAE architecture (Fig.~\ref{fig:vae_framework}) implements the variational autoencoder framework~\cite{kingma2022autoencodingvariationalbayes} to learn compressed representations of VQE circuit parameters. VAEs have been widely used for disentangled representation learning~\cite{kim2019disentanglingfactorising, chen2019isolatingsourcesdisentanglementvariational} and generative modeling across diverse domains~\cite{ho2020denoisingdiffusionprobabilisticmodels, rombach2022highresolutionimagesynthesislatent, radford2021learningtransferablevisualmodels}. The model processes a dataset $\{\boldsymbol{\theta}^{(j)}\}_{j=1}^M$ where each $\boldsymbol{\theta}^{(j)} \in \mathbb{R}^n$ represents VQE-optimized parameters for a Hamiltonian $H(\boldsymbol{x}^{(j)})$ at physical parameter $\boldsymbol{x}^{(j)}$.

\subsubsection*{Encoder Architecture}

The encoder maps high-dimensional input parameters to a low-dimensional latent distribution through a carefully designed pipeline. Each input vector $\boldsymbol{\theta} \in \mathbb{R}^n$ (where $n$ is the number of gate parameters in the ansatz circuit) first passes through a 1D convolutional neural network with three layers of increasing channel depth ($c_1 = 8$, $c_2 = 16$, $c_3 = 32$). The CNN serves two critical functions: first, it expands the one-dimensional parameter sequence into a multi-channel feature representation $\mathbf{X}_{\text{cnn}} \in \mathbb{R}^{n \times c_3}$, effectively increasing the representational capacity of the network; second, through the sliding window operation of convolution kernels, it captures local dependencies between adjacent parameters in the circuit. This local structure is physically meaningful because quantum gates that are adjacent in the ansatz circuit often represent correlated operations on nearby qubits.

The CNN features are then processed by a multi-head self-attention (MHSA) layer, which computes attention weights between all pairs of parameter positions in the sequence. Unlike the CNN, which is limited to local neighborhoods, the MHSA module can model dependencies between arbitrary parameters regardless of their positional separation~\cite{Chen2025quantum,Rocchia2024unveiling}. This capability is essential for capturing the non-local structure of quantum many-body states, where entanglement can exist between qubits that are far apart in the circuit representation. The MHSA outputs a context-aware feature map $\mathbf{X}_{\text{attn}} \in \mathbb{R}^{n \times c_3}$ where each position contains information aggregated from the entire sequence.

After the attention layer, the feature map is flattened and passed through a multi-layer perceptron (MLP) that outputs the parameters of the latent distribution: a mean vector $\boldsymbol{\mu} \in \mathbb{R}^{L_z}$ and a log-variance vector $\log\boldsymbol{\sigma}^2 \in \mathbb{R}^{L_z}$, where $L_z \ll n$ is the latent space dimension. These parameters define the approximate posterior distribution $q_\phi(\boldsymbol{z}|\boldsymbol{\theta}) = \mathcal{N}(\boldsymbol{\mu}, \text{diag}(\boldsymbol{\sigma}^2))$. Following the reparameterization trick~\cite{kingma2022autoencodingvariationalbayes}, we sample latent variables as $\boldsymbol{z} = \boldsymbol{\mu} + \boldsymbol{\sigma} \odot \boldsymbol{\epsilon}$ with $\boldsymbol{\epsilon} \sim \mathcal{N}(\boldsymbol{0}, \boldsymbol{I})$. This reparameterization allows gradients to backpropagate through the stochastic sampling operation during training, enabling end-to-end optimization via stochastic gradient descent.

\subsubsection*{Decoder Architecture}

The decoder mirrors the encoder's CNNs structure in reverse, implementing a deconvolutional architecture that reconstructs the original parameter vector from the latent representation. The latent vector $\boldsymbol{z} \in \mathbb{R}^{L_z}$ is first expanded through an MLP, then passed through transposed 1D-convolution layers that progressively decrease the channel dimensions ($c_3 = 32$, $c_2 = 16$, $c_1 = 8$) while increasing the sequence length back to $n$. The decoder outputs a reconstructed parameter vector $\boldsymbol{\hat{\theta}} \in \mathbb{R}^n$ that aims to match the original input $\boldsymbol{\theta}$.

The latent space serves a dual purpose: during training, it provides the compressed bottleneck that forces the network to learn efficient representations; during inference, the latent vectors $\boldsymbol{z}$ can be extracted and analyzed using classical methods such as PCA, GMM, or t-SNE to identify quantum phase boundaries without requiring access to the high-dimensional parameter space or the original quantum states.

\subsection{Loss Function}

The VAE is trained by minimizing the evidence lower bound (ELBO), which decomposes into a reconstruction term and a KL divergence regularization term:

\begin{equation}
    \mathcal{L}_{\text{VAE}} = \frac{1}{M} \sum_{j=1}^{M} \left[ \|\boldsymbol{\theta}^{(j)} - \boldsymbol{\hat{\theta}}^{(j)}\|^2 + \beta D_{\text{KL}}\big( q(\boldsymbol{z} | \boldsymbol{\theta}^{(j)}) \| p(\boldsymbol{z}) \big) \right]
    \label{eq:vae_loss}
\end{equation}

where $M$ is the number of samples and $L_{\text{in}}$ is the input dimension (number of circuit parameters). The hyperparameter $\beta$ controls the strength of the KL regularization term~\cite{chen2019isolatingsourcesdisentanglementvariational}.

We employ mean squared error (MSE) rather than cross-entropy for the reconstruction loss. While cross-entropy is appropriate for probabilistic data following a known distribution, MSE is statistically equivalent to learning the conditional expectation $\mathbb{E}[\boldsymbol{\theta} | \boldsymbol{z}]$. This choice is motivated by our observation that the average parameter configuration within each quantum phase exhibits statistical consistency despite numerical fluctuations, while the mean configuration undergoes qualitative changes across phase boundaries. By optimizing MSE, the VAE learns to encode these phase-dependent average configurations into the latent space, enabling unsupervised phase detection.

The KL divergence term is given by:

\begin{equation}
D_{\text{KL}}(q \| p) = \frac{1}{2} \sum_{i=1}^{k} \left( \sigma_i^2 + \mu_i^2 - \log \sigma_i^2 - 1 \right)
\end{equation}

where $\boldsymbol{\mu}$ and $\boldsymbol{\sigma}^2$ are the mean and variance of the posterior distribution $q(\boldsymbol{z} | \boldsymbol{\theta})$, and $k = L_z$ is the latent space dimension. This term constrains the posterior to approximate the standard normal prior $p(\boldsymbol{z}) = \mathcal{N}(\boldsymbol{0}, \boldsymbol{I})$.

When setting $\beta=0$, the VAE loss function contains only the reconstruction loss $\mathcal{L}_{\text{recon}}$, and the model degenerates into a regular AutoEncoder (AE), i.e., no regularization term is introduced and only the reconstruction loss is used to learn the low-dimensional representation of the data. We will see later that if generation capability is not considered and only low-dimensional representations of data are learned through reconstruction loss, AE combined with PCA or KPCA is sufficient in many cases, and even performs better.

\section{Results}
\label{sec:results}

To verify the effectiveness and universality of our method in recognizing quantum phase transitions (QPTs) and to elucidate the factors affecting the performance of our classical neural network, we conducted numerical experiments on multiple tasks.

\subsection{Transverse Field Ising Model and Generalized Order Parameter}

\begin{figure}[htpb]
    \centering
    \includegraphics[width=0.48\textwidth]{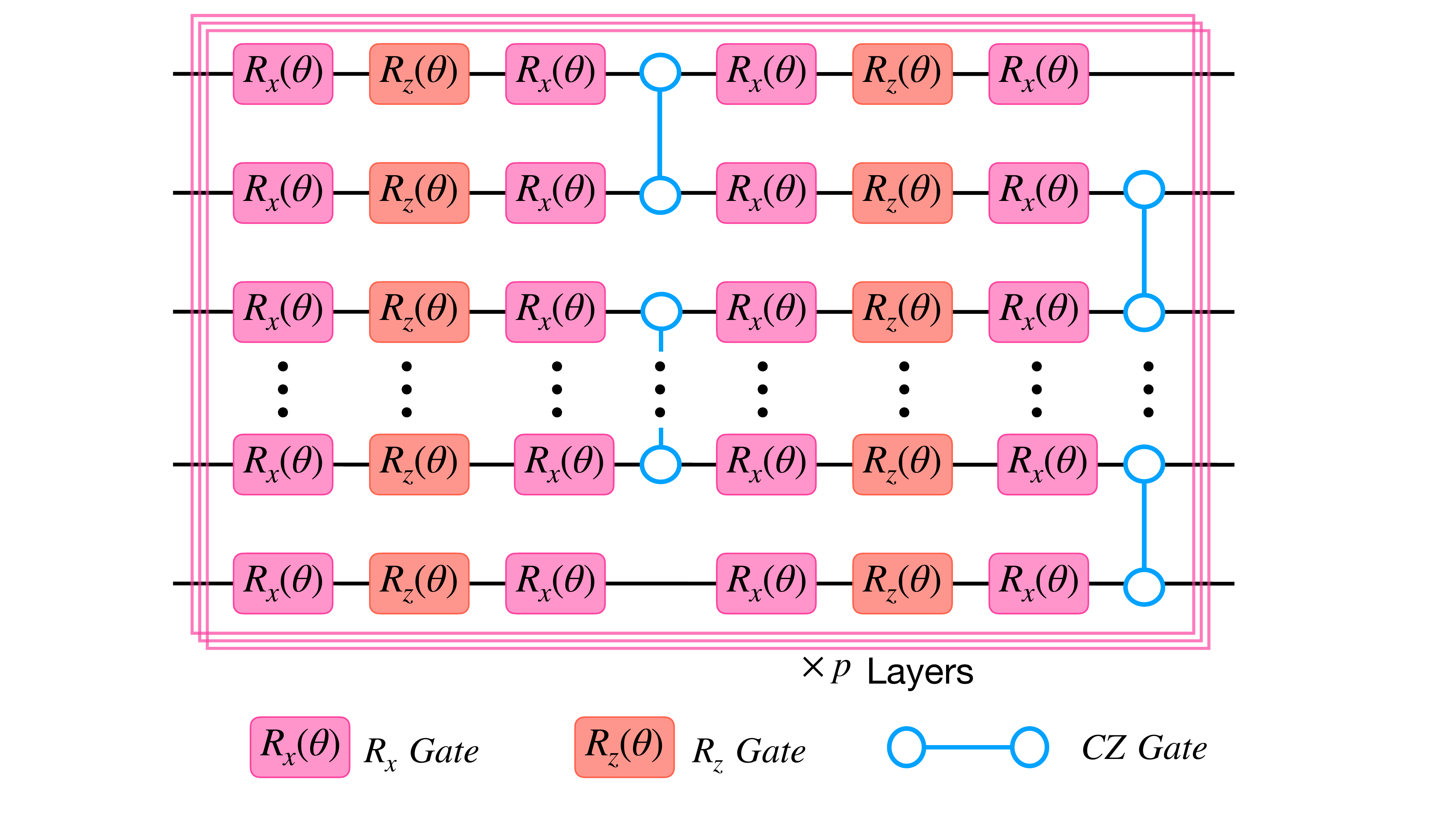}
    \caption{VQE ansatz circuit diagram for solving TFIM ground states under different physical parameters. The parameterized circuit consists of alternating single-qubit $R_x, R_z$ gates and parity-interleaved $\text{CZ}$ gates.}
    \label{fig:vae_ising_2design_circuit}
\end{figure}

We first applied our method to the task of learning quantum phase transitions in the Transverse-Field Ising Model (TFIM) described by the Hamiltonian 
\begin{equation}
    H = -J \sum_{i=1}^{N-1} Z_i Z_{i+1} - h \sum_{i=1}^N X_i.
\end{equation}
It is well-known that the TFIM undergoes a QPT at $h = 1$. For generality, we considered system sizes $N = 8, 12, 16$, and for each $N$, we performed VQE experiments with hardware-efficient ansatzes of $p = 4, 8, 12, 14, 16$ circuit layers, resulting in $3 \times 5 = 15$ datasets (tasks). For each task, we selected $h \in \{0, 0.001, ..., 2.0\}$ (2001 data points), with each $\boldsymbol{\theta}$ obtained via VQE with consistent initializations and iteration steps using the ansatz shown in Fig.~\ref{fig:vae_ising_2design_circuit}. As shown in the figure, each circuit layer contains $6N$ parameters, so the input vector length is $6Np$ depending on $N$ and $p$.

\begin{figure}[htpb]
    \centering
    \includegraphics[width=0.48\textwidth]{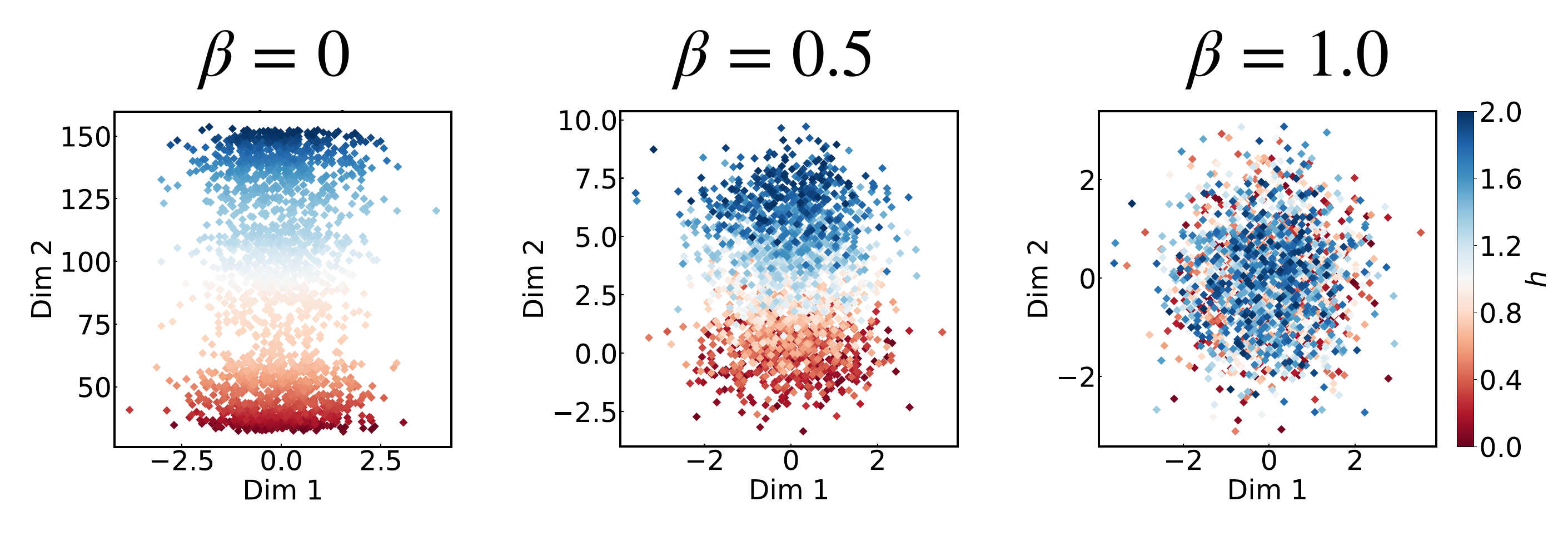}
    \caption{Distribution of 2001 TFIM samples with $N=12, p=14$ in the VAE latent space for $h \in [0, 2.0]$.}
    \label{fig:vae_ising_n12_p14_z2_b0-1_latnet}
\end{figure}

As an example, we first consider the task with $N=12, p=14$, yielding an input vector length of $6 \times 12 \times 14 = 1008$. We set the VAE latent space dimension $L_z=2$, $\beta=0.5$ in $\mathcal{L}_{\text{VAE}}$, and trained the model for 1000 epochs until convergence. The distribution of 2001 samples in the latent space is shown in Fig.~\ref{fig:vae_ising_n12_p14_z2_b0-1_latnet}(b). The data form two adjacent clusters, aggregating by the physical parameter $h$. With a boundary near 2.5 on the second latent dimension (vertical axis), samples with $h < 1$ cluster in the lower half (red) and those with $h > 1$ in the upper half (blue), with samples near $h \sim 1$ around the boundary. This demonstrates that our VAE model learns the QPT at $h=1$ and successfully distinguishes samples by their $h$ values.

To further investigate the influence of the regularization term (KL divergence) weight $\beta$ on model representation ability and verify the necessity of the current parameter setting, we compared latent space distributions for $\beta=0$ and $\beta=1$. When $\beta=0$, the model degenerates to a deterministic autoencoder with only reconstruction loss. As shown in Fig.~\ref{fig:vae_ising_n12_p14_z2_b0-1_latnet}(a), the vertical axis spans $0-150$, indicating severe geometric distortion and scale imbalance between the two dimensions due to the lack of KL divergence constraints on the latent space range and shape. In other high-dimensional mapping experiments (e.g., $L_z=8$), $\beta=0$ even causes samples to collapse into distorted, unclustered curves that are completely inseparable. Conversely, when $\beta=1$, the regularization term is too large, leading to posterior collapse where the latent space becomes a pure Gaussian distribution with samples from different physical parameters $h$ completely mixed, making them indistinguishable.

In our high-compression-ratio task, the reconstruction loss $\mathcal{L}_{\text{recon}}$ has a relatively small magnitude. A conventional $\beta=1$ would make the KL divergence term dominate the loss function, forcing the model to sacrifice reconstruction accuracy to meet the distribution constraints. Therefore, given the data characteristics of this task, selecting a smaller $\beta$ value (such as 0.5) is crucial for balancing reconstruction quality and latent space structure: it avoids both the geometric imbalance of $\beta=0$ and the feature annihilation of $\beta=1$.

An empirical guideline for choosing $\beta$: appropriately increase $\beta$ as the input dimension $L_{\text{in}}$ increases. This is because $L_{\text{in}}$ increases the magnitude of $\mathcal{L}_{\text{recon}}$, while the magnitude of the KL divergence term is independent of $L_{\text{in}}$. Similarly, appropriately decrease $L_z$ to keep $\beta$ larger. Preliminary small-scale calculations can estimate the magnitudes of the loss terms to determine a suitable initial $\beta$ value.

\begin{figure}[htpb]
    \centering
    \includegraphics[width=0.48\textwidth]{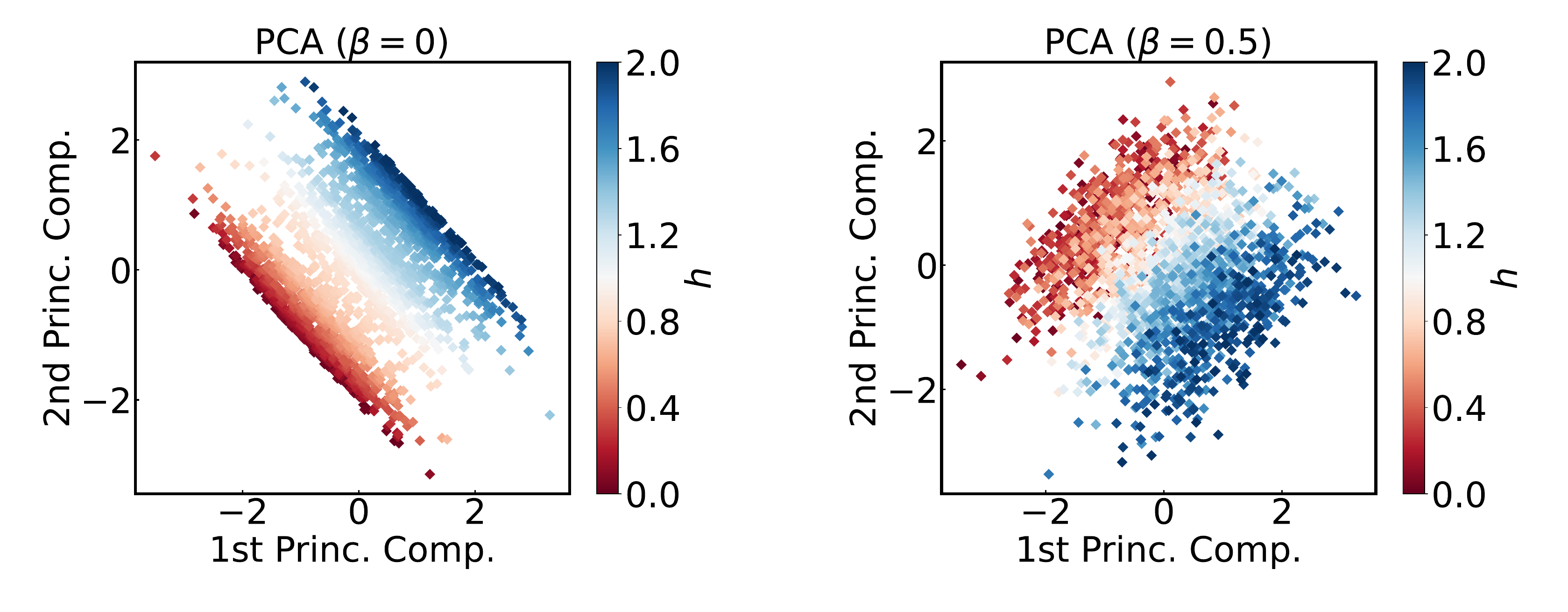}
    \caption{PCA results for latent space vectors $\mathbf{z}$ from TFIM with $N=12, p=14$.}
    \label{fig:vae_ising_p14_b_0N0.5_pca}
\end{figure}

Another interesting finding is that we can directly identify a "generalized order parameter" by performing PCA or KPCA on the latent variables. For the cases with $\beta=0$ and $\beta=0.5$, we performed PCA on the latent space vectors $\mathbf{z}$, as shown in Fig.~\ref{fig:vae_ising_p14_b_0N0.5_pca}. Since the original latent space is already 2D, PCA geometrically corresponds to an orthogonal rotation to find the "principal axis" with maximum variance.

The figure shows that data separation shifts from the vertical axis (Dimension 2) to the diagonal direction after PCA. This indicates that the "maximum variance" direction (PC1) is a linear combination of both latent dimensions, not a single original axis. This rotation "purifies" the data by extracting the direction of the most significant change driven by the phase transition parameter $h$, making the boundary between the ordered and disordered phases more pronounced. Standardization before PCA eliminates magnitude differences between dimensions, ensuring they contribute equally to principal component analysis, while mean-centering highlights fluctuations around the average state. PCA thus verifies the robustness of the VAE-learned manifold structure and reveals the physical pattern in a statistically significant way.

\begin{figure}[htpb]
    \centering
    \includegraphics[width=0.48\textwidth]{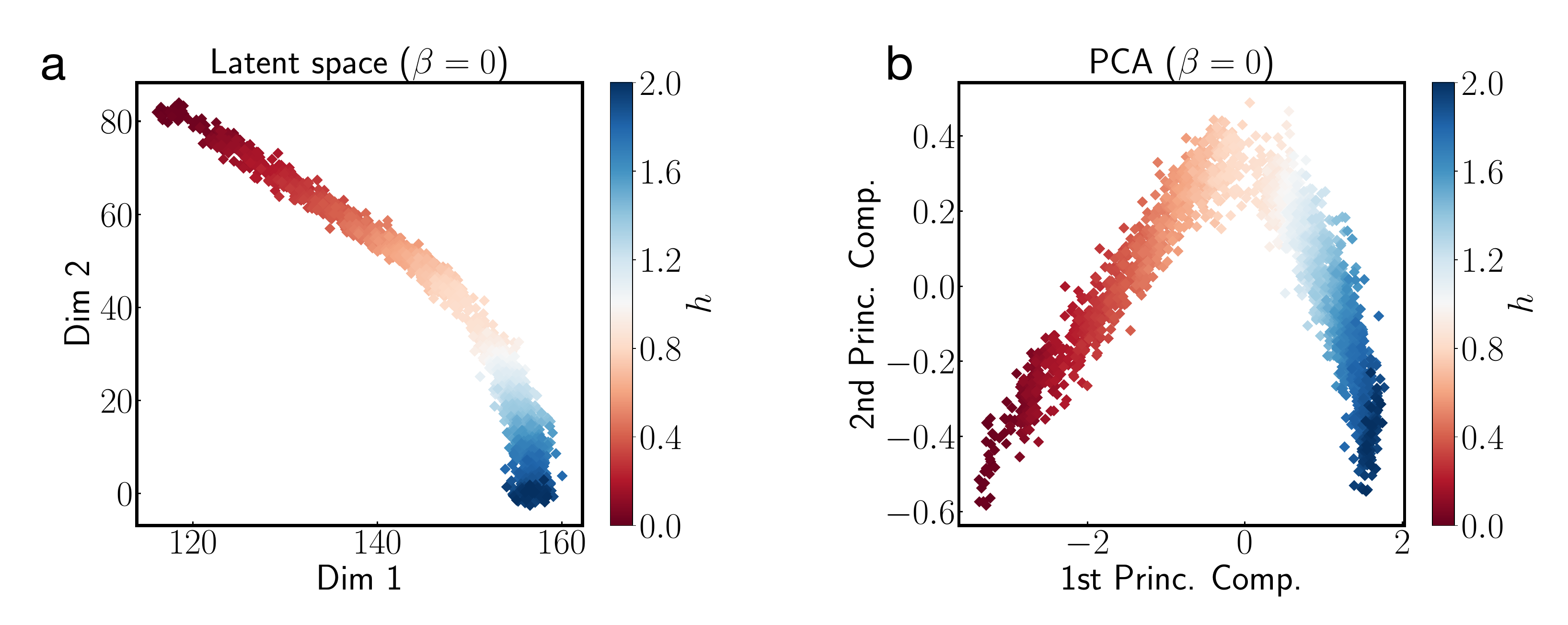}
    \caption{Distribution of 2001 samples for TFIM $N=12, p=8$ in (a) latent space and (b) after PCA processing.}
    \label{fig:vae_ising_p8_b0_l2_latNpca}
\end{figure}

A more extreme example is the task with $p=8, L_z = 2, \beta=0$. Here, the distribution of samples in the latent space is extremely anisotropic, making it impossible to directly discern phase-related distributions, as shown in Fig.~\ref{fig:vae_ising_p8_b0_l2_latNpca}(a). However, after correcting the coordinate system with PCA and removing the large offset, the originally tiny transverse structure is revealed. We can still see two distinct clusters separated by a clear boundary corresponding to the phase transition point, as shown in Fig.~\ref{fig:vae_ising_p8_b0_l2_latNpca}(b).

Comparing Fig.~\ref{fig:vae_ising_p8_b0_l2_latNpca} and Fig.~\ref{fig:vae_ising_p14_b_0N0.5_pca}, we observe a key phenomenon: although the latent space distribution at $\beta=0$ (deterministic AE) is far inferior to the standard normal distribution in terms of regularity, exhibiting significant anisotropy and irregular distortion, the clarity of the phase boundary after PCA projection is actually better than that of the well-regularized VAE.

\begin{figure}[htpb]
    \centering
    \includegraphics[width=0.38\textwidth]{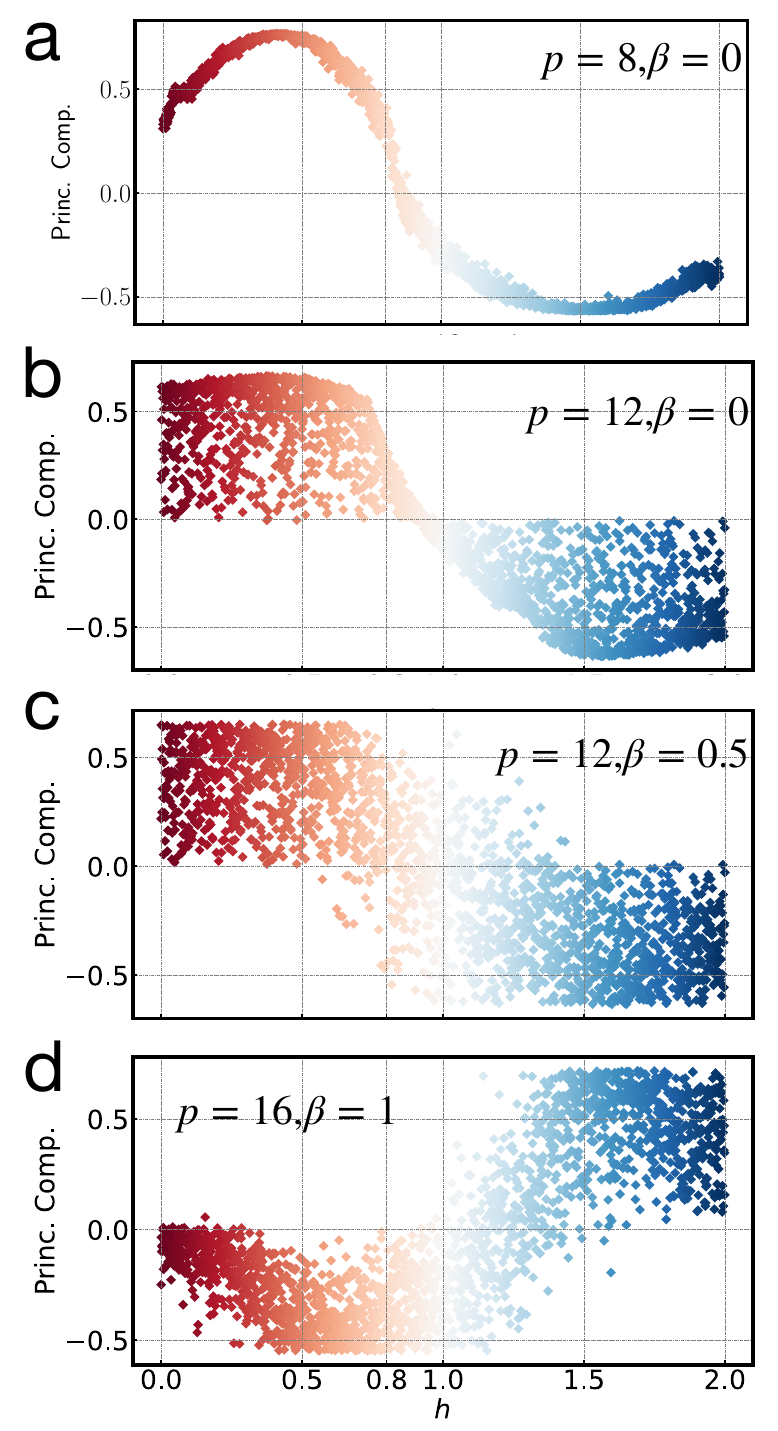}
    \caption{Kernel PCA results for TFIM showing clear phase separation near $h=1$. The manifold structure reveals the continuous evolution from ordered to disordered phase.}
    \label{fig:vae_ising_kpca_order}
\end{figure}

To further distill the physical meaning of this geometric feature, we introduced Kernel PCA (KPCA) on the basis of PCA linear dimensionality reduction to correct the nonlinear manifold of latent representations under different input complexities ($p$) and regularization constraints ($\beta, L_z$). We plotted the results for $p=8, \beta=0, L_z=2$; $p=14, \beta=0, L_z=2$; $p=14, \beta=0.5, L_z=2$; and $p=16, \beta=1, L_z=8$ in Fig.~\ref{fig:vae_ising_kpca_order}. At this point, the data manifold collapses to a single principal component, exhibiting strong physical regularity: this principal component uses 0 as a critical boundary to divide samples of different physical phases into positive and negative intervals (corresponding to $[0, 0.5]$ and $[-0.5, 0]$ respectively), while samples near the phase transition point smoothly cross the 0 point. Comparing Fig.~\ref{fig:vae_ising_kpca_order}(a)-(b) (AE) and (c)-(d) (VAE), we find that the feature distribution extracted by AE presents higher compactness and clearer separation; in contrast, VAE, due to the introduction of KL divergence noise, causes sample points to exhibit larger dispersion on the principal component axis, making the transition between phases blurrier.

It is also worth mentioning that since the two dimensions of the latent space are random, the direction of the principal component is also random. Being in the positive or negative half-region does not correspond to a specific quantum phase. Therefore, we can see that the direction in Fig.~\ref{fig:vae_ising_kpca_order}(d) is exactly opposite to that in (a)-(c).

This behavioral characteristic is highly consistent with the order parameter in physics. It is worth noting that since the system size is limited to $N=12$ lattice points, the real phase transition manifests as a transition in the region $h_x \in [0.8, 1.0]$ (finite-size effect~\cite{PhysRevB.89.094516}), rather than a perfect crossing of zero at $h=1$. Based on this, we regard this single principal component extracted by KPCA, which can quantitatively characterize phase separation and symmetry breaking, as a data-driven "generalized order parameter" for this system.

In the aforementioned KPCA analysis, we noticed that the VAE regularization mechanism causes latent space sample points to exhibit certain dispersion near the order parameter manifold. Although this high-frequency fluctuation reflects the randomness of generation, it masks the smooth trend of the order parameter evolving with the external field $h_x$. To extract deterministic physical laws from random fluctuations, we introduced a "sliding window mean strategy". Specifically, we defined a tiny neighborhood window on the parameter space of the external field $h_x$ and averaged the KPCA principal component values of all sample points falling within this window. This operation is mathematically equivalent to a local low-pass filter, which effectively filters out the random noise introduced by KL divergence. The mean behavior of the KPCA principal component can more accurately locate the system's phase transition point or the phase transition crossing region under finite size.

\begin{figure}[htpb]
    \centering
    \includegraphics[width=0.48\textwidth]{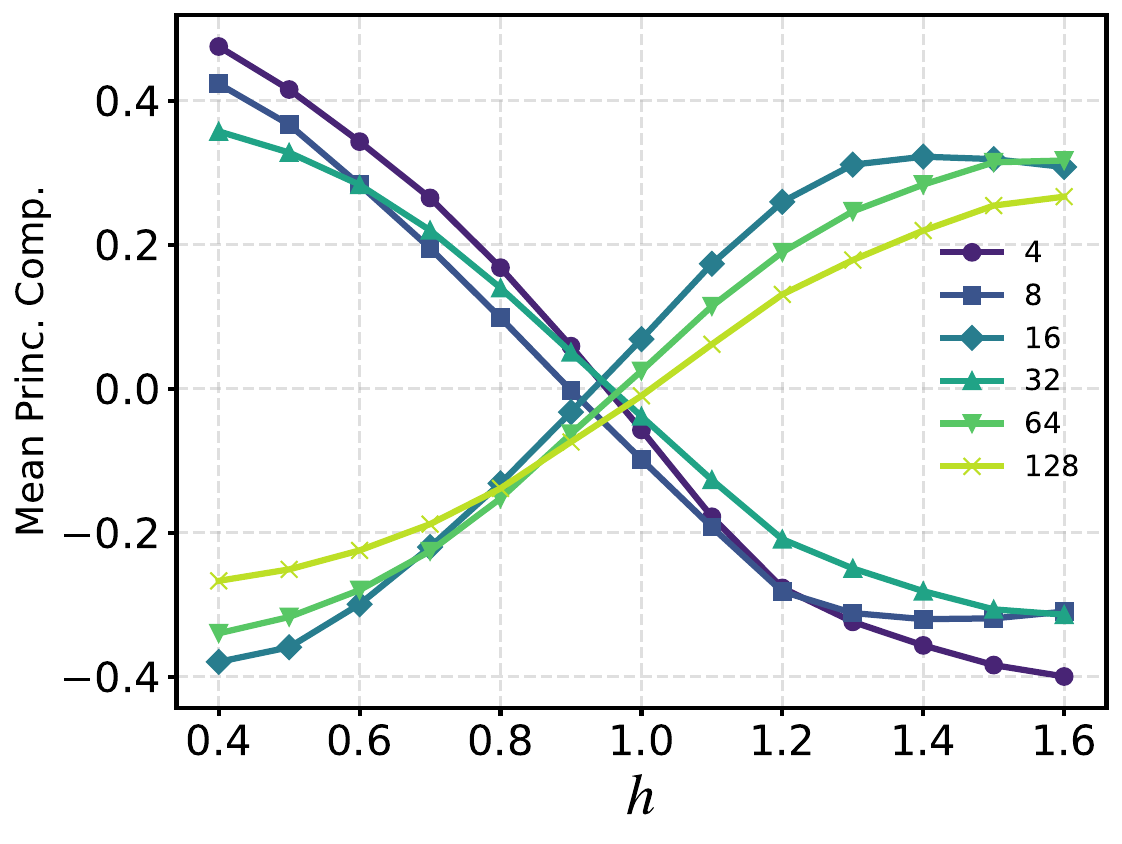}
    \caption{Generalized order parameter (first principal component) extracted from VAE latent space as a function of $h$ for TFIM. The sharp transition near $h=1$ clearly indicates the quantum phase transition point.}
    \label{fig:order_parameter_plot}
\end{figure}

Based on the above smoothing treatment, we further explored the influence of model hyperparameters (such as latent dimension $L_z$, regularization coefficient $\beta$, etc.) on the behavior of the order parameter. As shown in Fig.~\ref{fig:order_parameter_plot}, we calculated the generalized order parameter curves extracted under different VAE model configurations for the $N=16, p=12$ task and plotted them in the same coordinate system.

Since the direction of PCA/KPCA eigenvectors is arbitrary, some curves show an upward trend with increasing $h_x$, while others show a downward trend. This leads to an intersection in the region near $h_x \approx 1.0$. This phenomenon is similar to Finite-Size Scaling in statistical physics~\cite{PhysRevB.89.094516}: order parameter curves (or Binder cumulants) of different system sizes intersect at the critical point.

We regard the intersection point of various curves as a "model-independent fixed point". This phenomenon indicates that the position of the critical point has invariance to model hyperparameters (such as latent dimension or regularization strength). The intersection of all curves in the figure in the $(0.9, 1.0)$ interval effectively determines the phase transition point under the finite-size system. This not only validates the physical meaning of the "generalized order parameter" but also establishes a robust, data-driven localization method that does not depend on specific model parameters.

The role of PCA goes beyond this. For some tasks, the length $L_{\text{in}}$ of the input data $\boldsymbol{\theta}$ is too large or the internal structure is too complex, so that the effective information contained therein cannot be directly compressed by VAE or AE to an arbitrarily low-dimensional space (such as $L_z = 2$). In this case, PCA can project the original high-dimensional data to a two-dimensional space, retaining the main features of the data while removing noise and redundant information, to obtain a phase-related distribution.

For example, in the $N=16, p=12$ task mentioned above, the input data $\boldsymbol{\theta}$ has a length of $L_{\text{in}}=3Np = 6 \times 16 \times 12 = 1152$. The reason why we did not plot the case of $L_z=2$ in Fig.~\ref{fig:order_parameter_plot} is that if the data is compressed to 2 dimensions via VAE at this time, meaningful information cannot be obtained at all. However, if the data is compressed to $L_z \geq 4$ dimensions, and then the most important component is found via PCA, a distribution structure related to the phase can be seen.

We analyze that when $L_z$ is forcibly limited to 2, in order to minimize the global reconstruction error, the neural network is forced to perform a kind of "lossy compression", that is, it is mathematically impossible to establish a reasonable mapping from $\mathbb{R}^{1152}$ to $\mathbb{R}^2$. This forced compression leads to a severe dimensional collapse effect: trajectories that are originally distinctly different in physical properties (i.e., far apart in $\mathbb{R}^{1152}$) overlap in space when violently projected onto the $\mathbb{R}^2$ plane. This overlap is not simple noise, but aliasing of the manifold structure, causing the dimensions carrying key physical information to be submerged by compression noise, thereby making the latent space present a meaningless chaotic state and unable to carry out meaningful distribution.

Conversely, appropriately increasing the latent dimension to $L_z=4$ constructs a space sufficient to carry information. Although the 8-dimensional space at this time is still far smaller than the original 1152 dimensions, it provides enough geometric degrees of freedom for the encoder to "unfold" the curled manifold. At this time, the encoder can first complete the nonlinear disentanglement task, eliminate the self-intersection nodes of the manifold, and maintain the geometric separation of different physical phases. At this time, although the data is opaque to intuitive observation in the 8-dimensional space, its intrinsic evolution trajectory related to physical information has become a linearly separable structure.

On this basis, introducing PCA is not purely dimensionality reduction, but a principal axis correction of the disentangled manifold. The phase transition-related information extracted by the attention mechanism will manifest as the feature mode with the most significant information content and the most drastic change in this latent space. This physical signal, specifically extracted and enhanced by the model, is reflected in the statistical distribution as the direction of maximum variance, thus naturally corresponding to the principal component direction with the largest eigenvalue of the covariance matrix. Therefore, PCA can accurately identify and extract the main hyperplane representing phase transition evolution from the 8-dimensional unfolded manifold, filtering out redundant degrees of freedom unrelated to the physical process.

A more significant example on this point is the Cluster-Ising model learning task discussed in the next section.

\subsection{A More Complexity Task: Cluster-Ising Model}

To verify the universality of the VAE framework in more complex topological systems, we extend our study to the Cluster-Ising model:

\begin{equation}
H = -\sum_{i=1}^{N-2}Z_iX_{i+1}Z_{i+2} - h_1 \sum_{i=1}^{N}X_i - h_2 \sum_{i=1}^{N-1}X_iX_{i+1}
\label{eq:cluster_ising}
\end{equation}

Unlike the Ising model, this system has three quantum phases: paramagnetic phase (PM), Ising phase, and symmetry-protected topological (SPT) phase protected by $\mathbb{Z}_2 \times \mathbb{Z}_2$ symmetry. The SPT phase is driven by the interaction term $-\sum_{i=1}^{N-2}Z_iX_{i+1}Z_{i+2}$, and its ground state is often called a cluster state or graph state, which plays an important role in measurement-based quantum computing and quantum error correction.

\begin{figure}[htpb]
    \centering
    \includegraphics[width=0.48\textwidth]{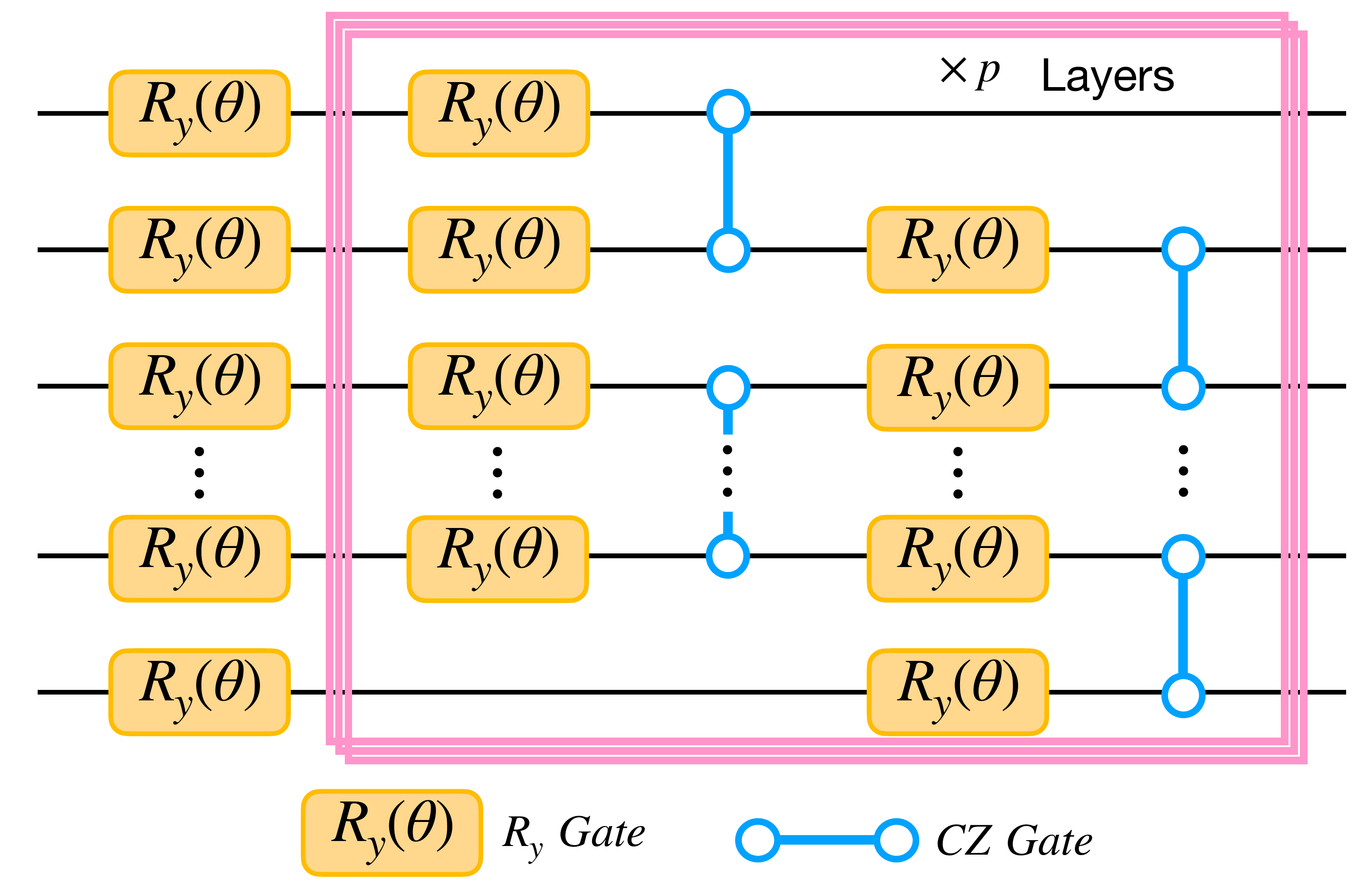}
    \caption{VQE ansatz circuit diagram for solving Cluster-Ising ground states under different physical parameters. The parameterized circuit consists of alternating single-qubit $R_y$ gates and parity-interleaved $\text{CZ}$ gates.}
    \label{fig:vae_cluster_2design_circuit}
\end{figure}

We demonstrate the framework with a 15-qubit system. We use the circuit shown in Fig.~\ref{fig:vae_cluster_2design_circuit} with $p=15$ layers. We sample $h_1 \in [0.1, 1.2]$ with step 0.1, and for each $h_1$, sample $h_2 \in [-2.3, 1.6]$ with step 0.001, generating 3901 samples per $h_1$ value. Following the methodology used for the Ising model, we take the VQE-converged circuit parameters $\boldsymbol{\theta}$ as raw data. According to the chosen circuit structure, the input data dimension is given by $L_{\text{in}} = (N-1)2p + N$, which in this case is $14 \times (2 \times 15) + 15 = 435$ dimensions. For each $h_1$ value, this scanning process generates 3901 sample points (i.e., 3901 vectors $\boldsymbol{\theta}$ of length 435) corresponding to different $h_2$. We input these datasets for specific $h_1$ into the VAE model separately, mapping them to a low-dimensional latent space through the encoder to extract latent vectors capable of characterizing the phase transition features of quantum states.

It is worth noting that although the input vector dimension (435 dimensions) here is smaller than the input scale in the Ising model task, this does not mean the problem is simpler. On the contrary, because the Cluster-Ising model has a more complex topological structure and multi-phase coexistence characteristics, the physical information density carried by its circuit parameters is significantly higher than that of the Ising model in the previous text. Therefore, we cannot simply compress the dimension directly to $L_z=2$; an overly low latent space dimension is insufficient to completely encode the rich phase transition information of this model. Instead, we need to set a higher latent space dimension in the VAE to ensure it contains phase-related information, and then subsequently process the data to 2 dimensions via PCA and standardization procedures for subsequent GMM and other operations.

\begin{figure}[htpb]
    \centering
    \includegraphics[width=0.48\textwidth]{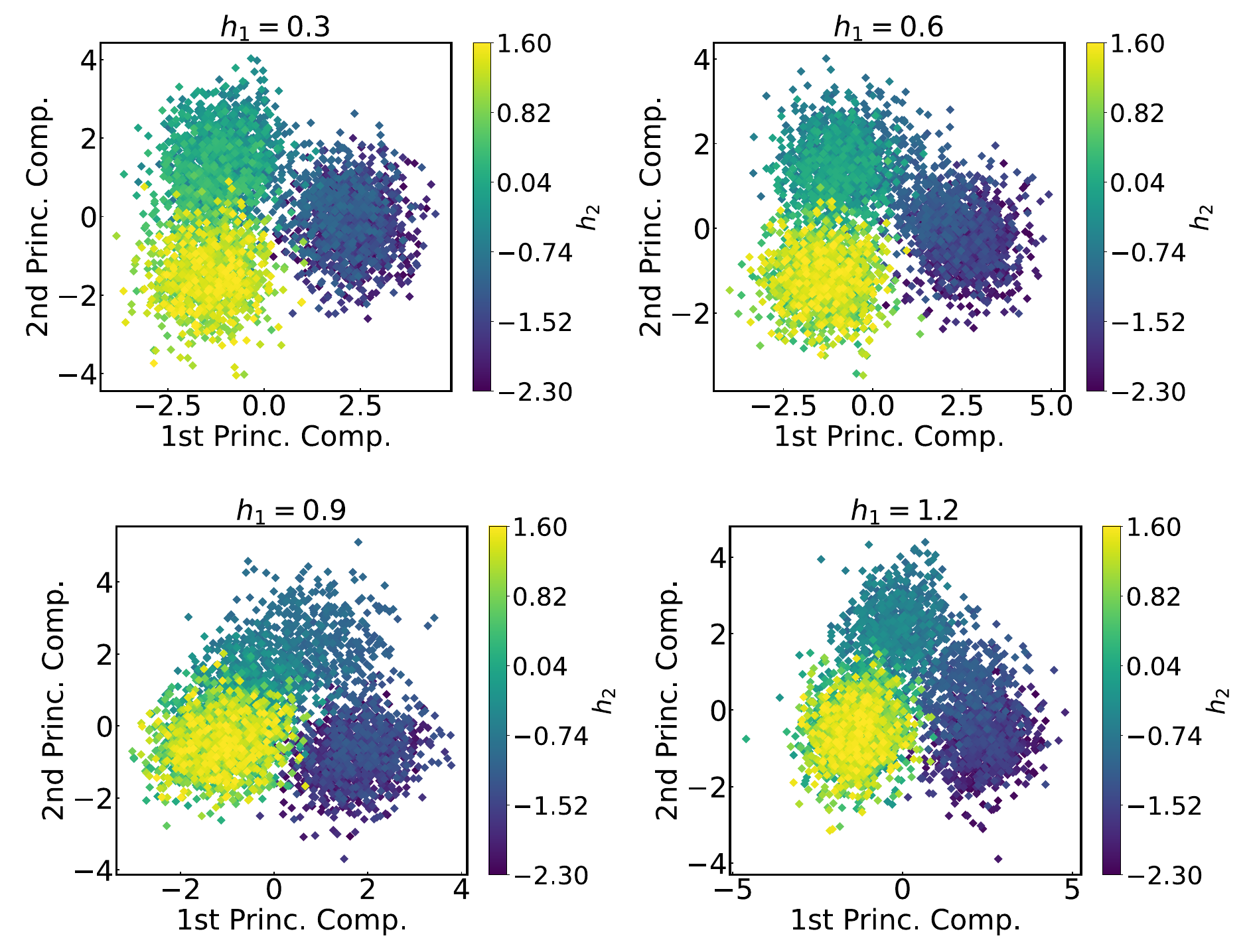}
    \caption{Distribution of Cluster-Ising model data in principal component space for $h_1 \in \{0.3, 0.6, 0.9, 1.2\}$ after VAE encoding to $L_z=16$.}
    \label{fig:cluster_ising_pca}
\end{figure}

After reducing dimensionality with VAE ($L_z=16$) and PCA projection, data shows distinct clustering structures in principal component space corresponding to different quantum phases, as shown in Fig.~\ref{fig:cluster_ising_pca}. Observing the figure, we can see that the data presents a significant clustering structure in the principal component space, and sample points corresponding to different quantum phases are effectively separated in the low-dimensional space. However, comparing datasets with different $h_1$ horizontally, we can find that their PCA projection effects vary: some datasets show a regular three-cluster structure with clear boundaries, while others are not so clear.

This morphological difference stems from two levels. From a physical perspective, the adjustment of $h_1$ changes the system's Hamiltonian, causing changes in the entanglement structure and physical information of the quantum state; the elements in the phase corresponding to each cluster would not be the same. From a data perspective, the statistical properties of different datasets are inherently different: parameter sets $\boldsymbol{\theta}$ corresponding to different $h_1$ have different data structures, including differences in numerical scale, dynamic range (extremum distribution), and statistical moments (mean and variance). Theoretically, if we fine-tune hyperparameters (such as adjusting latent space dimension $L_z$ or regularization coefficient $\beta$, etc.) for each dataset individually, we could indeed obtain low-dimensional representations with more perfect morphology.

Nevertheless, we did not pursue extreme tuning for each parameter point, because the subsequently introduced PCA and GMM have robustness. Experimental results show that GMM does not demand that the dimensionality reduction results of VAE must present "perfect isotropy" or "standard geometric distribution"; even when the data distribution is distorted to a certain extent, GMM can still perform accurate label classification based on data density. This characteristic not only avoids tedious point-by-point parameter fine-tuning but also strongly proves the universality and generalization ability of this analysis framework when facing different physical environments.

\begin{figure}[htpb]
    \centering
    \includegraphics[width=0.48\textwidth]{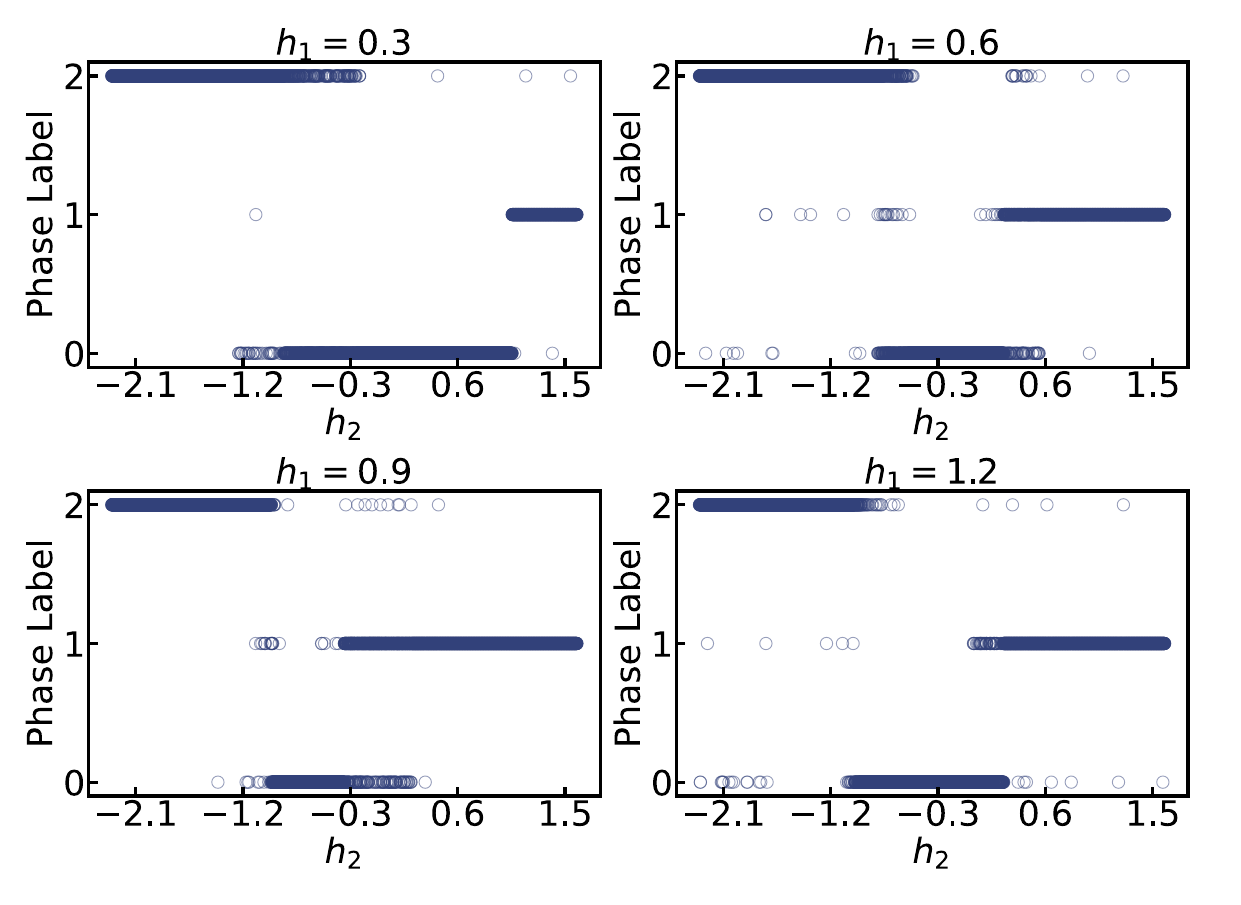}
    \caption{GMM cluster prediction results for Cluster Ising model with $h_1 \in \{0.3, 0.6, 0.9, 1.2\}$.}
    \label{fig:cluster_ising_gmm_label_addition}
\end{figure}

Employing GMM with $K=3$ components, we successfully classify the samples. Figure~\ref{fig:cluster_ising_gmm_label_addition} shows the GMM cluster labels for datasets with $h_1 \in \{0.3, 0.6, 0.9, 1.2\}$. As $h_2$ changes, labels form stable plateaus corresponding to different phases, with abrupt changes indicating phase transitions. It needs to be widely explained that due to the randomness of VAE latent space formation and GMM algorithm initialization, the labels $\{0, 1, 2\}$ output by GMM do not have absolute or fixed physical definitions. In other words, in one training session, label "0" may correspond to the SPT phase, while in another independent training session or analysis of another dataset, it may correspond to the Ising phase or PM phase. Therefore, these numerical labels are not direct names for physical phases, but markers for the intrinsic cluster structure of the data. Their main function lies in distinction—that is, marking out three regions with distinct statistical characteristics in the parameter space, thereby verifying the existence of three obvious phases in the system.

To intuitively display this clustering effect, we visualized the GMM clustering prediction results for four typical datasets with $h_1 \in \{0.3, 0.6, 0.9, 1.2\}$, as shown in Fig.~\ref{fig:cluster_ising_gmm_label_addition}. In this figure, the horizontal axis represents the scanning parameter $h_2$, and the vertical axis is the clustering label output by GMM (taking values $0, 1, 2$).

From the figure, we can observe that as $h_2$ changes in the range $[-2.3, 1.6]$, data points maintain the same label value within specific $h_2$ intervals, forming continuous plateau regions; between these stable regions, label values undergo abrupt changes, meaning that the cluster to which the sample points belong has switched. These mutation points of GMM classification labels under different parameters are exactly the phase transition points of the system. Therefore, we can find approximate phase transition points by locating the mutation points of GMM labels.

At the same time, we note that although the labels output by GMM have numerical randomness, this does not hinder their effectiveness as a data-driven generalized order parameter. In each independent dataset, we can observe that the label values $\{0, 1, 2\}$ form three stable plateaus, establishing a one-to-one correspondence with the three phases in the system. This characteristic provides distinct discrete characterization quantities for different quantum phases.

\begin{figure}[htpb]
    \centering
    \includegraphics[width=0.48\textwidth]{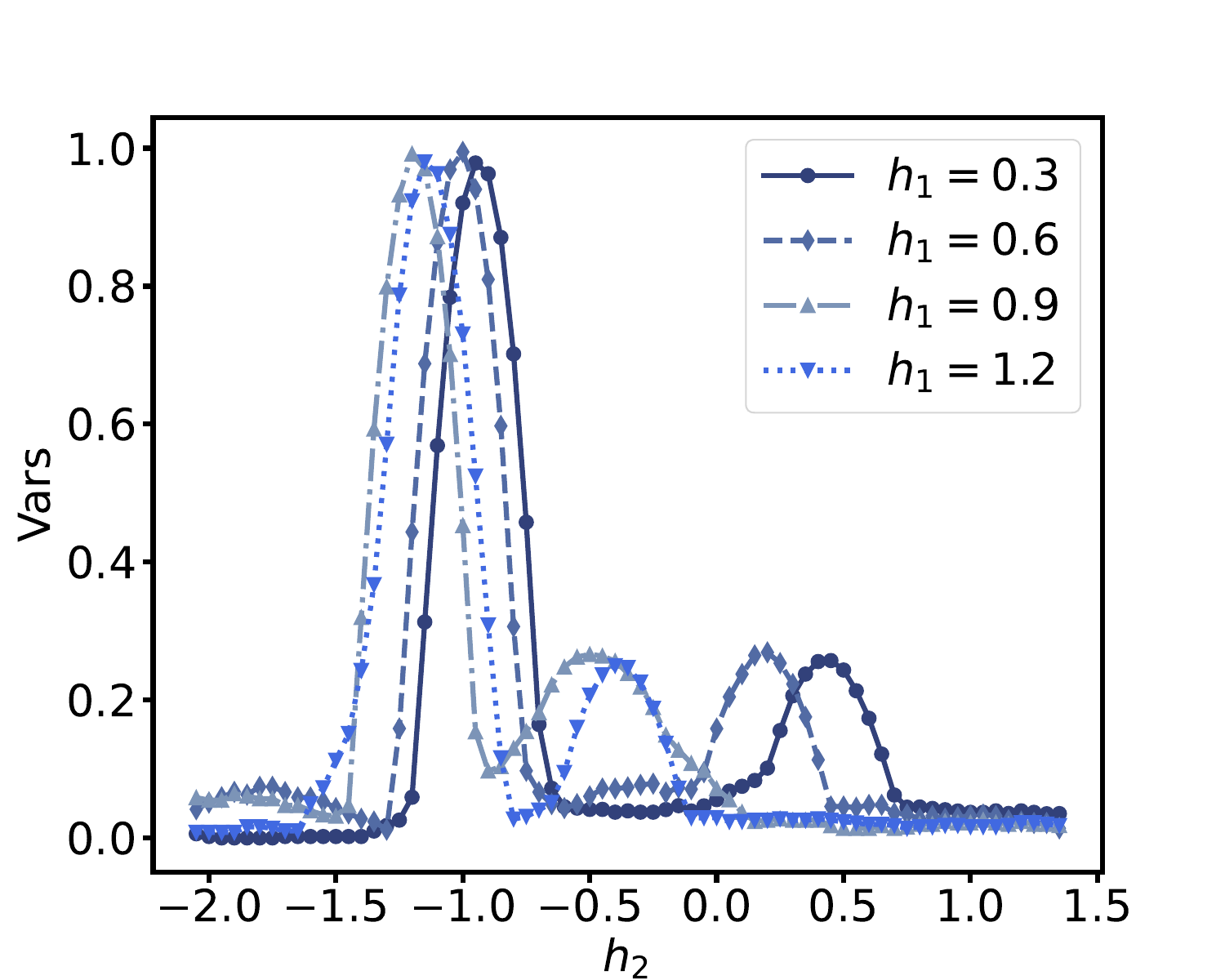}
    \caption{Local variance of GMM cluster labels as a function of $h_2$ for Cluster Ising model with $h_1 \in \{0.3, 0.6, 0.9, 1.2\}$.}
    \label{fig:cluster_gmm_vars_0.3_1.2}
\end{figure}

Using the sliding window variance method, we locate phase transition points by identifying regions with maximum variance in the GMM labels, as shown in Fig.~\ref{fig:cluster_gmm_vars_0.3_1.2}. The basis of this method is: in the phase-stable plateau region, the GMM label remains constant, and its local variance approaches zero; while in the critical neighborhood where phase transition occurs, the label value will switch violently between different categories (e.g., jumping from $1$ to $2$), causing the local variance to increase significantly. For our task, we calculate the variance through sliding window scanning, accurately identifying two maximum points in the variance curves corresponding to datasets with different $h_1$. These two peak positions correspond to the two phase transition points in the current parameter range. Fig.~\ref{fig:cluster_gmm_vars_0.3_1.2} shows the variance evolution curves for datasets corresponding to $h_1 \in \{0.3, 0.6, 0.9, 1.2\}$, clearly verifying the existence of the double-peak structure.

More importantly, this positioning scheme based on variance peaks provides a basis for conducting numerical quantitative finite-size scaling analysis: by systematically examining GMM clustering results under different lattice numbers $N$, VQE circuit depths $p$, and VAE model hyperparameters, and extracting corresponding variance peaks for comprehensive scaling analysis, we will be able to extend this framework to the in-depth study of system critical exponents and universality class behaviors.

\begin{figure}[htpb]
    \centering
    \includegraphics[width=0.48\textwidth]{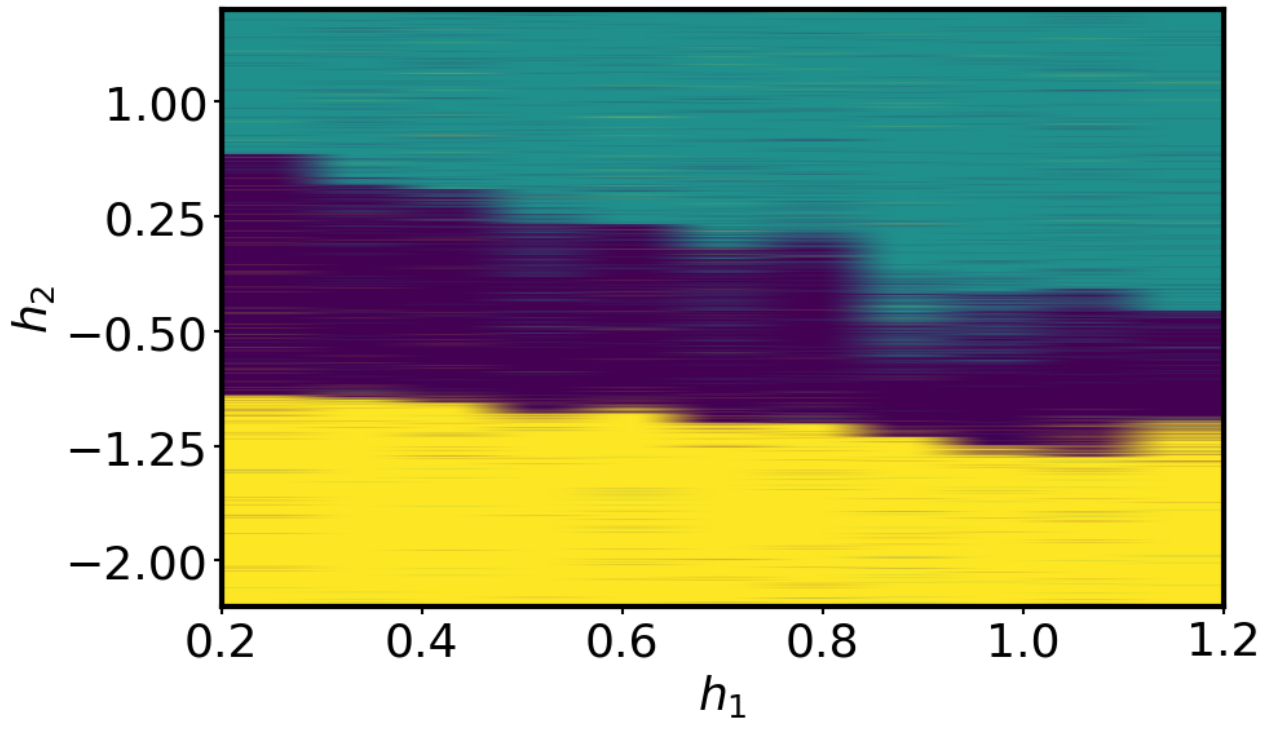}
    \caption{Complete phase diagram of the Cluster-Ising model in the ($h_1, h_2$) parameter space, reconstructed by aggregating phase transition points from all datasets.}
    \label{fig:cluster_pahse_diagram}
\end{figure}

This approach allows us to construct the complete phase diagram of the system across the entire physical parameter space by aggregating results for all $h_1$ values. We plotted the distribution of these labels on a two-dimensional plane with $h_1$ as the horizontal axis and $h_2$ as the vertical axis, successfully reconstructing the complete phase diagram of the model in the entire physical parameter space, as shown in Fig.~\ref{fig:cluster_pahse_diagram}.

\section{Discussion}
\label{sec:discussion}

\subsection{Analysis of Post-Processing Mechanism}

In the above research, we conducted extensive evaluations of the proposed variational quantum machine learning framework under various physical scenarios. The framework demonstrates robust quantum phase transition learning capabilities across different physical models (Ising model with symmetry-breaking phase transitions, Cluster-Ising model with SPT phase transitions), various VQE circuit configurations (such as different circuit depths $p$), and diverse VAE hyperparameter settings (such as $\beta$, $L_z$). Furthermore, by integrating different post-processing techniques—specifically including PCA for linear alignment and standardization, KPCA for nonlinear manifold correction, and GMM for probabilistic clustering—we successfully developed a generalized order parameter extraction method. These methods not only precisely locate physical phase transition points but also effectively reconstruct phase diagrams, thereby strongly demonstrating the effectiveness, robustness, and wide universality of this framework in handling quantum many-body problems.

Despite the above achievements, some concerns about the architecture remain, such as whether post-processing methods like PCA are absolutely necessary. Can we enable VAE to directly output disentangled order parameters in specific latent dimensions (e.g., $z_1$) through end-to-end training?

From the theoretical perspective of representation learning, building a "perfect VAE" that can directly output axis-aligned order parameters in specific latent dimensions is mathematically entirely feasible. This feasibility is mainly achieved by injecting strong physical inductive biases into the model, through specific approaches including anisotropic design of the prior distribution and explicit constraints on statistical independence.

First, directed encoding of physical features can be achieved by breaking the rotational symmetry of the latent space. A standard VAE typically assumes a latent prior of isotropic Gaussian distribution $p(\mathbf{z}) = \mathcal{N}(\mathbf{0}, \mathbf{I})$, whose KL divergence term $\mathcal{L}_{\text{KL}}$ has rotational invariance, making the model insensitive to the orientation of the latent manifold. However, if we introduce an asymmetric prior, such as setting $p(\mathbf{z}) = \mathcal{N}(\mathbf{0}, \boldsymbol{\Sigma}_{\text{prior}})$ where the diagonal covariance matrix $\boldsymbol{\Sigma}_{\text{prior}}$ has elements set to significantly different values (e.g., $\sigma_1^2 \gg 1$ for the dimension corresponding to the order parameter, and $\sigma_{j>1}^2 \ll 1$ for dimensions corresponding to noise), then the KL divergence term becomes a directional regularization operator. During optimization, to minimize the KL loss, the encoder will be forced to align the most rapidly changing, highest variance physical features from the input data to the specific dimension with the largest prior variance ($z_1$), thereby directly achieving principal component extraction and separation in end-to-end training.

Second, feature disentanglement can be enforced during training by introducing a decorrelation penalty in the loss function. Drawing on the design concepts of FactorVAE and $\beta$-TCVAE, we can add an explicit regularization term to the loss function $\mathcal{L}$, such as the sum of squares of the off-diagonal elements of the latent variable covariance matrix. This constraint is mathematically equivalent to requiring the neural network to dynamically find a set of basis during backpropagation such that the covariance matrix of the latent distribution is diagonalized. This mechanism essentially embeds the orthogonalization objective of PCA into the optimization objective of VAE, forcing the model to learn latent representations that are statistically independent of each other, and thus making each output dimension naturally correspond to the main modes of variation in the data (i.e., separation of order parameters and noise).

In this study, we chose a general VAE/AE architecture that is "imperfect" in a certain sense. We acknowledge that without specific physical constraints, the latent space output by such a general variational model often exhibits rotation, translation, or even distribution distortion of the coordinate system.

However, this framework precisely demonstrates its universality. We only require the VAE to complete the most difficult and general task—topological manifold learning, i.e., preserving the relative distances and continuity relationships between physical states—without enforcing automatic alignment with the physical coordinate system. On this basis, the introduction of post-processing methods such as PCA, KPCA, and GMM actually constructs a set of "general correction interfaces": the VAE/AE is responsible for mining the information related to phase transitions in high-dimensional data; PCA/KPCA is responsible for correcting and aligning the mined manifold to a physically interpretable or clearer coordinate system. This design ensures that the framework can handle different types of quantum phase transition tasks with a unified architecture, without the need to redesign the network structure or finely tune hyperparameters for each specific Hamiltonian.

\subsection{Quantum Correlation Extraction and Implicit Physical Information Mining Based on Attention Mechanism}

Another question arises: Is the attention mechanism necessary? Can we successfully extract quantum correlations and mine implicit physical information without introducing the attention mechanism? At least within our framework structure, we tried performing simple PCA/KPCA, fully connected layers, or 1D-CNN directly on the raw high-dimensional data $\boldsymbol{\theta}$, and found that they were all unable to directly extract features related to physical phase transitions. This aligns with findings from quantum machine learning literature, where classical neural networks have shown limited success in capturing complex quantum correlations without appropriate architectural inductive biases~\cite{schuld2019machine, YaoFramework2019}.

We must first recognize that the parameter vector $\boldsymbol{\theta}$ output by VQE is fundamentally different from traditional image or time-series data. In quantum many-body physics, the phase transition behavior of the system, especially topological quantum phase transitions, is determined by the global topological properties of the wave function. Moreover, our circuit parameters themselves are not quantum states, but a highly abstract "contextual representation" of the quantum gate operation sequence and its interaction environment. The proximity of parameter vector indices does not represent physical correlation; on the contrary, the change of the 1st parameter in the vector may be closely coupled with the 100th parameter, and both jointly determine the global physical state of the system. This long-range non-local dependency buried deep in the high-dimensional parameter space makes CNNs relying on local convolution kernels or PCA limited by linear assumptions unable to capture critical physical entanglement information.

We speculate that the core advantage of introducing the attention mechanism in the VAE architecture of this study lies in its ability to dynamically calculate the correlation weights between any two positions in the input sequence, constructing a fully connected global receptive field.

At the physical level, the attention matrix actually acts as a "quantum correlation detector". Through MHSA calculation, the model can automatically identify and focus on key parameter pairs that play a decisive role in phase transitions, regardless of how far apart they are physically in the input vector. This mechanism in a sense simulates the long-range entanglement effect in quantum many-body systems and extracts it from the circuit parameters, concentrating attention weights on those non-local parameter correlations containing physical information. It is this "semantic aggregation" capability that enables the VAE to strip hidden physical laws from messy raw parameters during the dimensionality compression process.

PCA, as a linear algorithm, relies entirely on the linear separability of input features. For raw high-dimensional PQC parameters, physical information is entangled in a highly nonlinear way, and direct application of PCA can only extract meaningless statistical variance and cannot separate physical phases. The attention layer first completes the most difficult nonlinear feature decoupling inside the VAE, extracting physical information hidden in complex correlations and mapping it to the latent space to form a manifold with a clear topological structure. The subsequent PCA is merely finding the best observation angle on this manifold that has been organized by attention. In other words, without the deep mining and reorganization of implicit physical information by the attention mechanism, the latent space would be chaotic, and subsequent PCA would be unable to extract clear principal components corresponding to phase transitions. Therefore, our architecture is a process where the attention mechanism mines physical correlations, VAE compresses the manifold, and PCA (or other post-processing means) extracts the main modes.

\begin{figure}[htbp]
    \centering
    \includegraphics[width=0.48\textwidth]{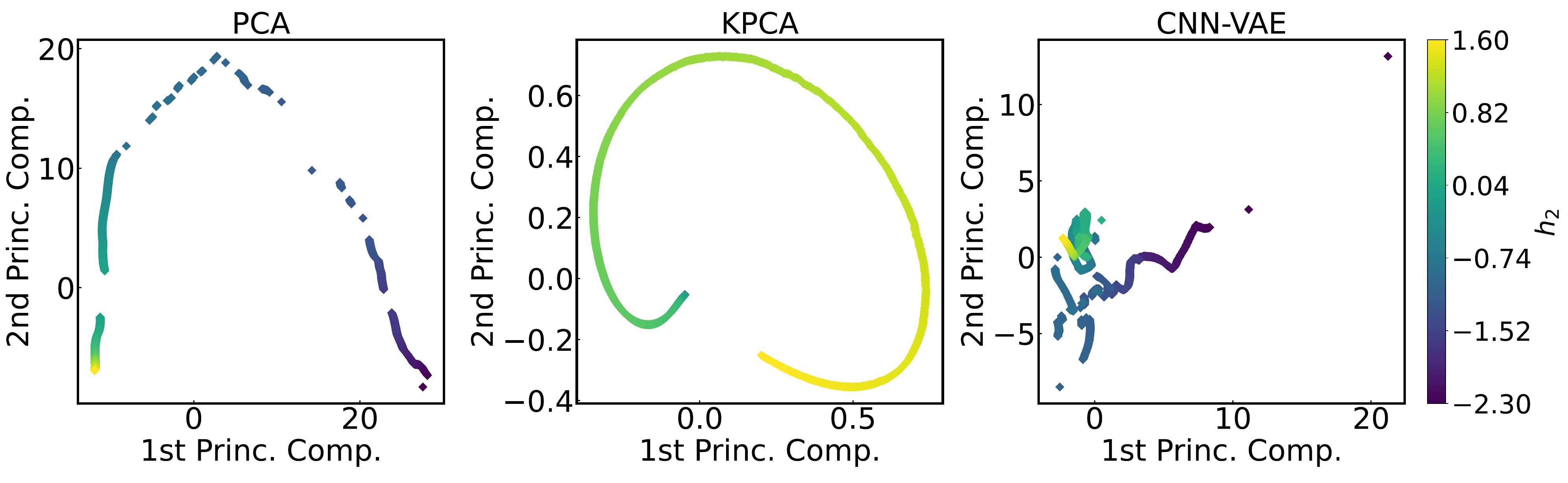}
    \caption{Schematic diagram of dimensionality reduction of raw data by different feature extraction methods. From left to right: direct PCA, direct KPCA, and distribution compressed by CNN-VAE. The raw data corresponding to this figure is the dataset of the 15-qubit Cluster-Ising model at $h_1=0.6$. The performance in other models and datasets is similar to this figure.}
    \label{fig:cvae_need_attention}
\end{figure}

To illustrate our analysis, we constructed a CNN-VAE framework that removes the attention mechanism and retains only the CNN neural network (replacing the original attention layer with CNN), and compared its dimensionality reduction effect on raw data with the results of direct PCA and KPCA dimensionality reduction feature extraction on raw data, as shown in Fig.~\ref{fig:cvae_need_attention}. From left to right: PCA, KPCA, and CNN-VAE dimensionality reduction results.

For PCA, the data points appear as a segmented, discontinuous inverted V-shaped trajectory in the 2D projection, with uneven data distribution composed of several dense line segments separated by obvious gaps. This indicates that PCA only captures the dominant explicit changes with the largest variance in the data—that is, the drastic numerical drift of the converged circuit parameters $\boldsymbol{\theta}^*$ driven by the continuous scanning of the Hamiltonian parameter $h_2$. This low-order, global change driven by the external field occupies an absolute dominant position in statistical variance, while high-order many-body correlations related to quantum phases are discarded due to weak variance contributions, resulting in circuit parameters containing rich quantum phase-related information being compressed into low-dimensional trajectories that only reflect parameter changes.

Unlike the discontinuous distribution of PCA, the data projected by KPCA forms a highly continuous and smooth arched manifold without obvious breakpoints or cluster separation. KPCA captures the nonlinear dependencies of data in high-dimensional space by introducing nonlinear kernel functions, smoothing the broken line trajectory in PCA. However, this smooth geometric structure indicates that the algorithm mainly retains the continuity characteristics of the quantum state evolution with the parameter $h_2$. Quantum phase transitions essentially involve symmetry breaking or changes in topological order, usually accompanied by singularities at critical points. For circuit parameters, this singularity may be hidden in deep data correlations. The KPCA results show that kernel methods based on global similarity metrics mask this singularity near the phase transition point, modeling the singular phase transition process as a smooth evolution of parameters, and thus cannot achieve corresponding data separation based on differences.

The latent space distribution of CNN-VAE presents a twisted and messy state, with scattered data points forming neither a continuous manifold nor clear clusters. This shows that even with a deep neural network structure with certain local perception capabilities, in the absence of an attention layer, the inductive bias of CNN does not match the deep correlations between many-body circuit parameters and the global physical characteristics of quantum phase transitions. CNNs are based on local receptive fields and pooling operations, tending to extract local texture features. In deep networks, this long-range parity correlation is gradually diluted or destroyed, leading to the network's inability to extract effective features capable of distinguishing different topological phases, thereby causing the latent space structure to collapse.

The above analysis shows that to accurately identify quantum phases, the model must have the ability to break local limitations. The attention mechanism with a global receptive field can indeed directly capture the hidden weak correlations between any two circuit parameters inside the input data $\boldsymbol{\theta}$, thereby extracting implicit phase-related information from the strong background noise dominated by a single variable, and even capturing data-driven generalized order parameters to achieve true phase classification and characterization.

\subsection{Evaluating VAE Performance via Quantum State Generation}

Although we have pointed out that as a general unsupervised learning framework, the VAE in this study may not automatically achieve perfect orthogonal decoupling in its latent space without specific physical constraints, this does not imply that the representation capability of the VAE itself is insufficient. On the contrary, in terms of manifold construction and information compression quality, the VAE framework has demonstrated excellent performance in capturing the topological structure of quantum many-body systems.

To further confirm this, i.e., to verify that the VAE architecture effectively encodes complete phase transition information in the latent space and constructs a high-quality data distribution, we introduced a Conditional Diffusion Model (CDM) based on the latent space for independent experimental evaluation. As a generative model, the conditional diffusion model can generate quantum states satisfying specific conditions in the latent space~\cite{rombach2022highresolutionimagesynthesislatent}.

Specifically, we trained a lightweight conditional diffusion model composed of fully connected layers in the latent space, enabling it to generate corresponding latent representations based on specified phase labels. Subsequently, we input these generated latent variables into the pre-trained decoder to map them into new parameterized circuit vectors $\boldsymbol{\theta}_{\text{new}}$, and reloaded these vectors into the ansatz circuit to prepare quantum states. Finally, by calculating the order parameter of the quantum states, we verified whether the generated quantum states faithfully preserved the physical characteristics of the target phase.

The core logic of this experimental design is that the generation quality of the diffusion model essentially depends on the distribution quality and structural clarity of the input data. If the VAE encoder fails to construct a latent space containing phase transition-related physical information, the subsequent diffusion model will be unable to capture effective probability distributions, and thus unable to generate quantum states carrying correct phase information. Therefore, this experiment confirms the VAE latent space's ability to learn phase transition-related information by examining the generation effect of the diffusion model.

The specific evaluation platform is a simplified Cluster-Ising model with $N=11$ qubits, with the Hamiltonian:

\begin{equation}
    H(\lambda) = \sum_{i=1}^N(X_{i-1}Z_iX_{i+1} + \lambda Y_iY_{i+1})
\end{equation}

Compared to Eq.~\ref{eq:cluster_ising}, the current model lacks one external field term and has one fewer parameter. According to theoretical research, this model has only two quantum phases: the SPT phase and the Antiferromagnetic (AFM) phase. These two phases can be characterized by the following order parameter~\cite{Smacchia_2011}:

\begin{equation}
    \mathcal{S} = \langle X_1X_2\left(\prod_{i=3}^{N-2}Z_i\right)Y_{N-1}X_N\rangle
\end{equation}

When $\lambda < 1$, the system is in the SPT phase, and the ground state has a finite $\mathcal{S} \in (0, 1]$. When $\lambda > 1$, the system is in the AFM phase, and the ground state has $\mathcal{S} = 0$.

We sampled from $\lambda \in [0, 0.8]$ (SPT phase) and $\lambda \in [1.2, 2.0]$ (AFM phase) with a step size of 0.001 to obtain all VQE-converged circuit parameter vectors $\boldsymbol{\theta}$ to constitute our dataset.

\begin{figure}[htbp]
    \centering
    \includegraphics[width=0.48\textwidth]{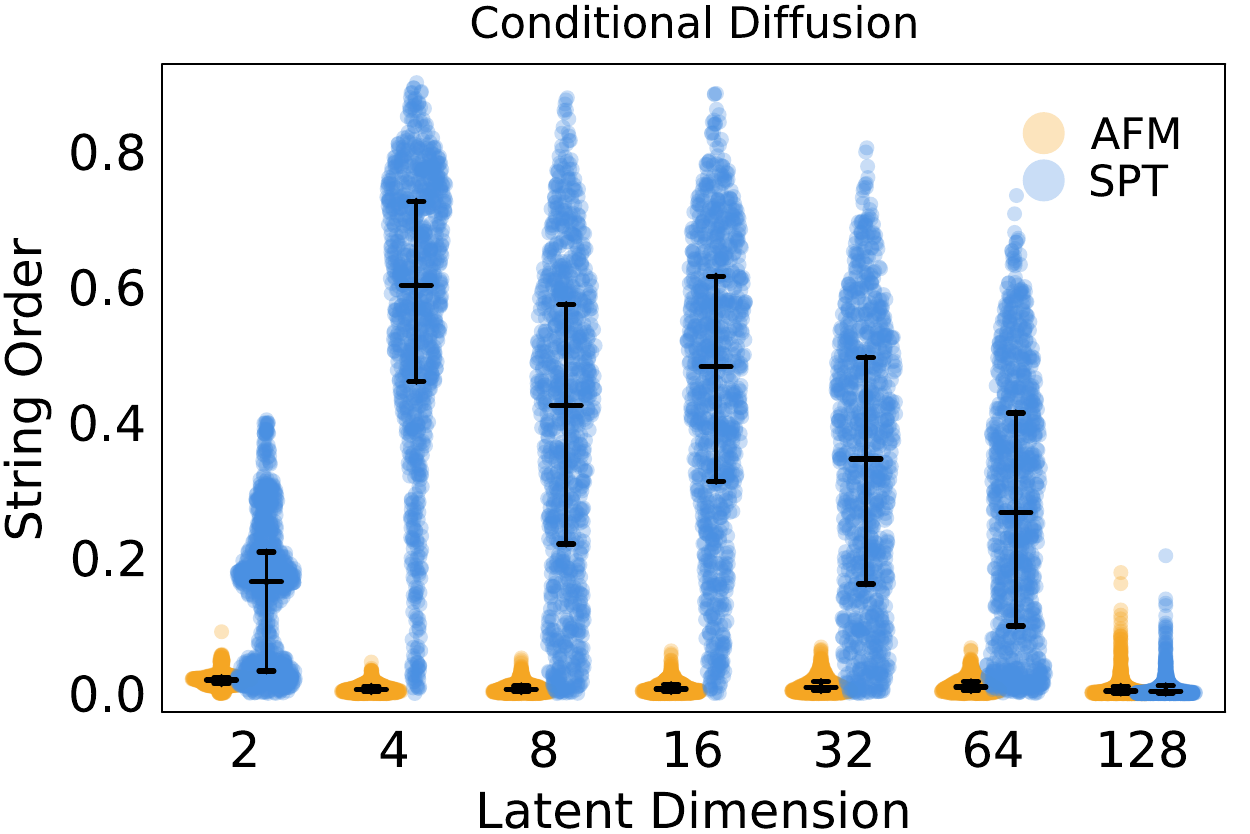}
    \caption{Distribution of order parameters for quantum states generated conditioned on SPT and AFM phases in VAE latent spaces of different dimensions. From left to right: $L_z = 2, 4, 8, 16, 32, 64, 128$. Blue and yellow dots represent SPT phase and AFM phase, respectively.}
    \label{fig:cdm_generated_s}
\end{figure}

We trained Conditional Diffusion Models (CDM) in VAE latent spaces of different dimensions ($L_z \in \{2, 4, 8, 16, 32, 64, 128\}$). To verify the generation performance, for each $L_z$ setting, we used the CDM to generate 100 samples conditioned on the SPT phase and 100 samples conditioned on the AFM phase, respectively. These generated latent vectors were decoded into new circuit parameters $\boldsymbol{\theta}_{\text{new}}$ and sent into the ansatz circuit to prepare quantum states, and finally, their corresponding order parameters $\mathcal{S}$ were calculated. The distribution of order parameters corresponding to these newly generated samples is shown in Fig.~\ref{fig:cdm_generated_s}.

At $L_z = 2$, whether generating quantum states conditioned on SPT or AFM, the order parameter $\mathcal{S}$ is concentrated in the low-value region of $<0.2$. This indicates that the latent space did not learn the phase-related information well, and the two-dimensional latent space did not cluster well according to phase information. This result confirms the previous conclusion: extremely low dimensions are insufficient to carry information describing such complex quantum phase transitions.

When $L_z = 4, 8, 16$, the quality of generated quantum states is the best. For the AFM condition, the generated state order parameter $\mathcal{S}$ is mainly distributed near 0, consistent with theoretical expectations; for the SPT condition, the order parameter is distributed in the $(0, 1]$ interval and significantly concentrated at higher values, which also matches theoretical expectations.

Starting from $L_z = 32$, although the generated states still generally follow the theoretically expected order parameter value distribution, the quality begins to decline compared to before, manifested by a decrease in the reachable maximum value of the order parameter and a diverging distribution. By $L_z = 128$, the generated states completely lose the physical meaning of the order parameter, with the order parameters of states generated under both conditions concentrated near 0.

We analyze that this is because starting from $L_z \sim 32$, the dimension of the latent space is excessive relative to the information contained within the input data $\boldsymbol{\theta}$. In high-dimensional cases, the KL divergence regularization of VAE makes it difficult to force the encoded posterior distribution to perfectly fill the entire high-dimensional Gaussian sphere. This causes the diffusion model to easily wander into "wasteland areas" that have never been mapped by the encoder during the early sampling stage. Since the core physical characteristics of the Cluster-Ising model are essentially determined by a small number of intrinsic degrees of freedom, an overly high latent space dimension (128 dimensions) not only fails to introduce more effective information but introduces a large number of noise dimensions. The diffusion model struggles to separate weak physical signals from these dominant noise dimensions during the denoising process, ultimately leading to the loss of key string order correlations in the decoded quantum states.

In summary, we verified through the conditional diffusion model that the VAE architecture not only effectively carries complete phase transition information in the latent space but also constructs a high-quality data distribution. This experimental result further establishes the effectiveness of our VAE architecture as a universal framework, proving its ability to learn and represent physical information closely related to quantum phases from VQE ansatz circuit parameters. At the same time, this conditional diffusion model also provides an effective evaluation method to help us confirm the optimal latent space compression dimension $L_z$ for specific physical models and input data. Finally, the conditional diffusion model in the latent space itself constitutes an efficient quantum state generation tool, capable of controllably generating new quantum states unseen under the target quantum phase.

\subsection{VQE Ansatz and Local Minima}

Quantum phases are physical characteristics of the system's ground state. Therefore, in general research, studying quantum phase transitions using VQE must rely on the precise preparation of the system's ground state. However, in this section, we will clarify that even if the VQE-converged parameter $\boldsymbol{\theta}$ is only a local minimum rather than the true ground state, our framework can still effectively identify phase transitions.

The core reason is that our model does not learn a specific quantum state in isolation, but learns the mapping distribution $P(\boldsymbol{\theta}^* | \boldsymbol{x})$ from the physical parameter space to the circuit parameter space under the same initialization strategy during the variational optimization process. Here, $\boldsymbol{\theta}^*$ represents the parameter configuration after VQE convergence (which may be the ground state or a local minimum). Although the quantum phase transition is defined by the mutation of the ground state properties of the Hamiltonian $H(\boldsymbol{x})$, this physical-level mutation will inevitably project onto the VQE optimization landscape.

As the external physical parameter $\boldsymbol{x}$ evolves, the change in Hamiltonian $H(\boldsymbol{x})$ causes its corresponding optimization landscape to undergo continuous or abrupt topological deformations. This deformation directly determines the convergence trajectory and final landing point of the classical optimizer. Systems within the same physical phase often share similar landscape geometric features (such as gradient flow directions and distribution positions of local minima), leading to a statistical "common clustering" of VQE convergence parameters $\boldsymbol{\theta}^*$. What our VAE mines is precisely this common distribution pattern driven by physical parameters $\boldsymbol{\theta}^*$ and hidden in convergence behaviors, which is independent of whether the converged state is the ground state.

We can intuitively explain this principle through an extreme example: suppose there is a physical system that has a relatively smooth convex optimization landscape in Phase A, where VQE easily converges to the true ground state; while in Phase B, due to the complexity of many-body entanglement, the optimization landscape becomes extremely rugged, causing VQE to easily fall into specific local minimum traps and fail to reach the ground state. In this case, although $\boldsymbol{\theta}^*$ in Phase B is not the true ground state, the two distinct optimization outcomes of "able to converge" and "falling into a trap" themselves are the strongest distinguishing features between Phase A and Phase B. We actually only need to learn whether a certain converged state is the ground state to determine which phase it belongs to. Although the real situation is more complex, our model learns and distinguishes quantum phases following such a principle.

\begin{figure}
    \centering
    \includegraphics[width=0.48\textwidth]{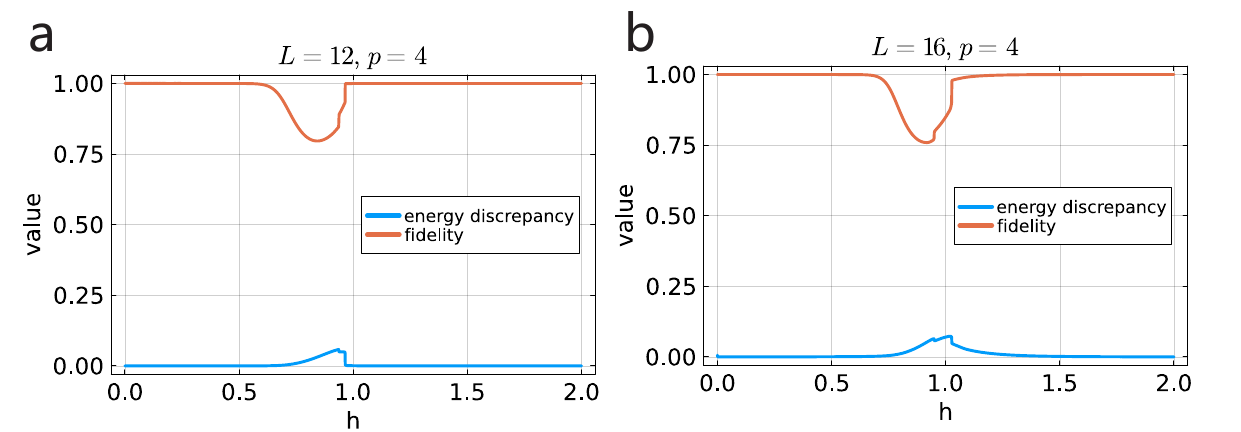}
    \caption{Energy difference and fidelity between VQE-converged states and true ground states for Ising model. Left: $L=12, p=4$; Right: $L=16, p=4$.}
    \label{fig:fe_ising}
\end{figure}

We provide numerical evidence supporting this argument: for an Ising model with system sizes $L=12$ and $L=16$ and circuit depth $p=4$, VQE converges to local minima in the region $h\in(0.5, 1.0)$ (see Fig.~\ref{fig:fe_ising}), yet our framework still successfully identifies phase transitions with high accuracy.

A more extreme example: for the 15-qubit Cluster Ising model, many samples have fidelity close to 0 and large energy differences, indicating convergence to excited states orthogonal to the ground state. However, our framework still successfully completes phase classification and phase diagram reconstruction, further demonstrating its exceptional robustness.

\begin{figure}
    \centering
    \includegraphics[width=0.48\textwidth]{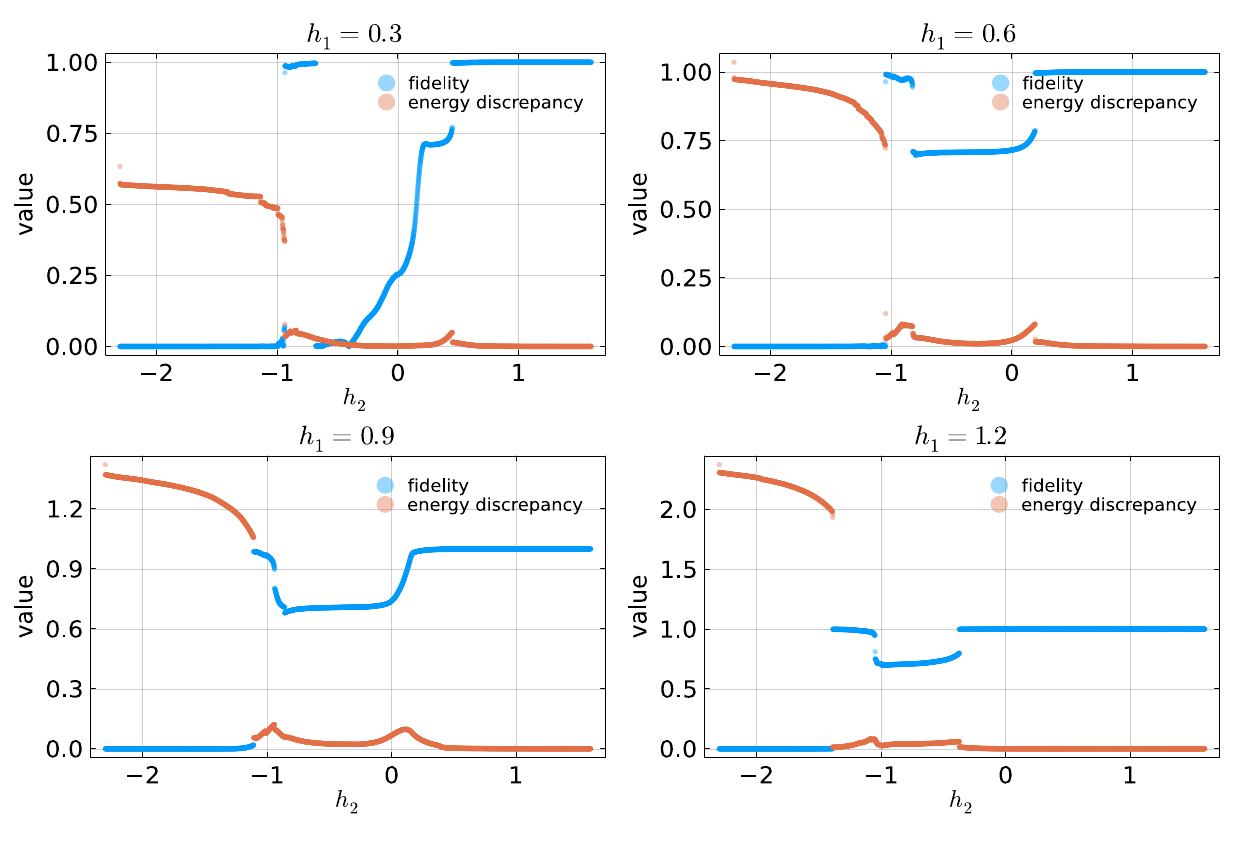}
    \caption{Fidelity and clustering results for VQE-converged states. For the Ising model with $L=8$, $p=4$, $b=0$, the results show clear division between high-fidelity states and low-fidelity states based on our clustering method.}
    \label{fig:fe_clustering_sm}
\end{figure}

To quantify the robustness to VQE local minima, we performed the following experiment. For the Ising model with $L=8$, $p=4$, $b=0$, we obtained a set of VQE-converged states with fidelity ranging from 0 to 1 and then clustered them using our framework. As shown in Fig.~\ref{fig:fe_clustering_sm}, the results indicate a clear division between high-fidelity states and low-fidelity states based on our clustering method. The fidelity difference between the two types of states is significant, confirming our framework's ability to distinguish different phases even when VQE converges to local minima.

Based on this characteristic, our model exhibits strong robustness when facing local minima, no longer demanding that VQE must strictly converge to the ground state. This characteristic effectively broadens the application boundary of VQE, greatly relaxing the constraints on its convergence performance, thus providing a more universal solution for studying quantum many-body problems under complex energy landscapes on NISQ devices.

\section{Conclusion}
\label{sec:conclusion}

In this paper, we propose a novel approach that shifts the research focus to VQE circuit parameters themselves for quantum phase transition (QPT) detection. Unlike traditional methods that rely on quantum state density matrices or observable measurements, our framework exploits correlations hidden in variational parameter space, opening a new path for processing quantum information in the NISQ era.

We present a classical AI framework integrating attention mechanism and variational autoencoder (VAE), systematically verifying its exceptional universality and robustness. Experimental evidence demonstrates that across multiple physical systems (including Ising models and Cluster Ising models), different circuit depths, and various hyperparameter configurations, this framework can consistently extract a "generalized order parameter" corresponding to physical phase transitions and reconstruct phase diagrams in physical parameter space. Additionally, leveraging the generative nature of VAE, the framework exhibits the capability to generate parameters for new, unseen quantum states, offering possibilities for numerical simulation and extended study of quantum states.

Most importantly, we deeply investigate and confirm that our framework demonstrates strong robustness to VQE local minima. Numerical analyses show that even when VQE fails to converge to the true ground state in the optimization landscape—even falling into excited states orthogonal to the ground state—the framework can still complete phase identification by capturing the distributional characteristics of converged parameters. This finding holds significant practical implications: our framework does not strictly require global optimal parameters or true ground states for QPT detection, thereby significantly relaxing the stringent requirements on VQE convergence performance.

In summary, the proposed framework provides a theoretical framework for QPT exploration and engineering improvements for the application potential of variational quantum algorithms in the NISQ era. By reducing dependence on high-precision parameter convergence, our framework shows high compatibility with current quantum hardware environments, offering a more practical solution for conducting complex many-body physics research on real, noisy quantum devices.

\begin{acknowledgments}
We thank Zhaohui Wei and Shiwei Zhang for helpful discussions. Quantum-related computations are performed using Yao.jl~\cite{YaoFramework2019}, and we also provide  MindQuantum~\cite{xu2024mindspore} implementations for certain quantum circuits, while PyTorch~\cite{paszke2019pytorchimperativestylehighperformance} is used for implementing and training the neural network models. This work is supported by 
the Beijing Institute of Technology Research Fund Program under Grant No. 2024CX01015, and the CPS-Yangtze Delta Region IndustrialInnovation Center of Quantum and Information Technology-MindSpore Quantum Open Fund.

\end{acknowledgments}

\section*{Code and Data availability}
The source code and data for this study are publicly available on GitHub: \url{https://github.com/zipeilee/attention-pqc}.

\bibliography{refs}

\end{document}